\documentclass[%
 preprint, 
 amsmath,amssymb,
 aps, physrev,
]{revtex4-2}

\usepackage{amsmath}
\usepackage{amssymb}
\usepackage{graphicx}
\usepackage{dcolumn}
\usepackage{bm}
\usepackage{tikz}
\usepackage{pgfplots}
\pgfplotsset{compat=1.18}
\usepackage{subcaption}

\begin{document}

\preprint{Accepted in CQG}

\title{\textbf{On Super-Hyperbolic Wormhole Geometries and their Deformations} 
}%

\author{Subrat Parida}%
 \email{Corresponding Author: paridasubrat553@gmail.com \\ The author is currently looking for doctoral position.}{}
\affiliation{%
 Department of Mathematics, Ramanujan School of Mathematical Sciences, Pondicherry University, Pondicherry, India.
}%


\begin{abstract}
In this study, we developed a unified geometric and relativistic framework for a new family of traversable wormholes constructed from a super-hyperbolic deformation of the classical catenoid, whose geometric embedding combines an explicit local curvature parameter $\kappa$, which secures a regular throat, with the Mittag-Leffler-based super-hyperboloid deformation function ${\mathfrak{S}C}_{\alpha} (v)$, where the deformation parameter $\alpha>1$ governs the radial onset and far-field growth of the embedding beyond the throat, continuously approaching the classical catenoidal geometry in the boundary limit $\alpha\rightarrow1^{+}$. Allowing a non-constant redshift $\Phi(r)$, we systematically evaluate closed analytical expressions for the shape function, curvature tensors, matter sources, and junction conditions, culminating in an exact thin-shell equation of motion and stability criterion. This work establishes a mathematically consistent and physically viable framework for the extension of wormhole theory, providing a new avenue for exploring exotic geometries within general relativity and identifying the conditions under which traversable configurations remain locally stable.  
\end{abstract}


\maketitle


\section{\label{sec:level1}Introduction}
\subsection{\label{subsec:level1.1}Background and Motivation}

Among the profound consequences of Einstein's general theory of relativity \cite{einstein1915feldgleichungen}, perhaps the most fascinating geometric intuition is that spacetime curvature can support non-trivial topologies \cite{thorne2014black}. In the decades that followed, many exact solutions \cite{schwarzschild1916gravitationsfeld, weyl1917gravitationstheorie, levi1917ds2, lewis1932some, andress1930some, papapetrou1953rotationssymmetrische, kruskal1960maximal, Szekeres:1960gm, kerr1963gravitational, boyer1967maximal, tomimatsu1972new, tomimatsu1973new} of Einstein Field Equations (EFEs) have been extensively studied. The first non-trivial solution is the Schwarzschild solution \cite{schwarzschild1916gravitationsfeld}, which describes the spacetime geometry outside a spherically symmetric non-rotating mass. The Schwarzschild solution laid the foundation for the modern understanding of black holes \cite{chandrasekhar1998mathematical, hawking2023large, thorne2014black}. Further developments led to more general black hole solutions, including the Kerr \cite{kerr1963gravitational} and Reissner-Nordström metrics \cite{reissner1916eigengravitation}.

In fact, since the early days of the theory, physicists and mathematicians have speculated about the possibility of tunnels connecting regions of spacetime at large distances - what is usually called a \emph{wormhole} \cite{hawking1988wormholes, morris1988wormholes, visser1995lorentzian, kar1996evolving}. The concept arises naturally within the framework of General Relativity (GR) \cite{birkhoff1923relativity, einstein1935particle}, which revolutionized our understanding of gravity by introducing the notion of curved spacetime geometry \cite{hawking2023large, thorne2000gravitation}. Wormholes are configurations that connect spacetime through their geometrical curvature \cite{nandi1997brans} rather than through the embedding of matter-energy in a flat Euclidean space \cite{morris1988wormholes, hawking1988wormholes, kar1996evolving, friedman1993topological, visser1995lorentzian}. Traditionally, the intuitive idea behind a wormhole is an embedding of two asymptotically flat universes \cite{thorne2000gravitation} connected by a smooth \emph{``throat"} surface of minimal area, much like a \emph{catenoid} - the classical minimal surface of revolution \cite{choudhury2025catenoid} discovered independently by Euler (1744) \cite{euler1744methodus} and Lagrange (1760) \cite{lagrange1761essai}.

The catenoid is a unique minimal surface of revolution \cite{osserman2013survey, abbena2017modern} with zero mean curvature, obtained by rotating a catenary about its direction. Its geometrical simplicity has made it a natural visual and mathematical analog of the wormhole throat \cite{visser2003traversable}; which describes a smooth, axially symmetric surface \cite{bohmer2007conformally} that ``flares out" from a minimal radius and returns asymptotically to flatness \cite{morris1988wormholes, visser1995lorentzian, hawking2023large}.

Early models of wormholes, initially introduced as the Einstein-Rosen bridges \cite{einstein1935particle}, were highly unstable. Interest in wormholes was renewed with the introduction of exotic matter, partly proposed \cite{lobo2003linearized, lobo2005energy, VisserHochberg1998PRL, hochberg1998null} to account for the observed late-time acceleration of the universe. Subsequently, it was shown that such structures could achieve stability \cite{poisson1995thin, eiroa2004linearized, eiroa2008stability, lobo2003linearized, bohmer2007conformally} in the presence of exotic matter that violates the Null Energy Condition (NEC) \cite{hochberg1998null, lobo2005phantom, lobo2005energy}. Morris and Thorne (1988) \cite{morris1988wormholes} adopted precisely this geometric intuition in their classical model of a static traversable wormhole \cite{visser2003traversable}. In the Morris--Thorne (MT) framework, the spacetime metric is conventionally written as
\begin{equation}\label{1.1}
ds^{2}
=
-
e^{2\Phi(r)}dt^{2}
+
\frac{dr^{2}}{1-\frac{b(r)}{r}}
+
r^{2}d\Omega^{2},
\end{equation}
where
\[
d\Omega^{2}
=
d\theta^{2}
+
\sin^{2}\theta\, d\phi^{2}.
\]
Here, the function $b(r)$ denotes the \emph{shape function}, which determines the spatial structure of the wormhole geometry, while $\Phi(r)$ represents the \emph{redshift function}, governing the gravitational redshift and tidal properties of the spacetime \cite{morris1988wormholes, visser1995lorentzian, choudhury2025catenoid}. A particularly important subclass corresponds to the case $\Phi(r)=\mathrm{constant}$, for which tidal forces vanish for static observers traversing the throat. Under this simplifying assumption, the embedding geometry acquires a direct correspondence with the classical catenoidal structure, with the choice
\[
b(r)=\frac{r_{0}^{2}}{r}
\]
representing one of the simplest embedded wormhole configurations. Although the MT construction established the theoretical possibility of traversable wormholes, the requirement of NEC-violating matter and the restriction to constant-redshift geometries impose significant physical and geometrical limitations.

In subsequent decades, a significant amount of effort was devoted to the generalization of wormhole geometries \cite{eiroa2004linearized, eiroa2004cylindrical, lobo2003linearized, eiroa2009thin, nandi1997brans, bohmer2007conformally, rahaman2006thin}, either by allowing the redshift function to vary or by embedding it in modified gravitational theories, such as scalar-tensor, Gauss-Bonnet, or f(R) and f(Q) gravity models \cite{accetta1990wormholes, bronnikov2009once, bronnikov2010notes, maeda2008static, kanti2012stable, jusufi2020wormholes, godani2022stability, panyasiripan2024charged, lobo2009wormhole, mazharimousavi2016wormhole, harko2013modified, hassan2021traversable, banerjee2021wormhole, parsaei2022wormhole}. Even for these generalizations, though, integer-order geometric functions $\cosh(v)$ and $\sinh(v)$ were generally used for embedding profiles. These latter encode smooth but integer-order curvature, with the geometry transitioning between the throat and asymptotic regions with a curvature scaling controlled by integer-order derivatives.
This naturally raises the question of whether wormhole geometries may admit more general curvature structures associated with fractional or non-local geometric behavior near the throat region.

Fractional calculus \cite{hilfer2000applications, Kilbas2006} indeed offers a framework in which this question can be naturally answered. In fact, it generalizes differentiation and integration to non-integer orders and hence provides the proper language for describing processes and geometries with long-range correlations, non-local interactions, or scale-dependent curvature smoothening. Within the last two decades, fractional differential geometry \cite{lazopoulos2016fractional}, as well as quite prominently, the employment of \emph{Mittag-Leffler (ML) functions}, have found groundbreaking applications in anomalous diffusion \cite{luchko2010fractional} and viscoelastic systems \cite{Mainardi2010} and even in some formulations of non-local gravity \cite{calcagni2010fractal, tarasov2008fractional}.

The ML function,
\begin{equation}\label{spg1}
E_{\alpha}(z)
=
\sum_{n=0}^{\infty}
\frac{z^{n}}{\Gamma(\alpha n+1)},
\qquad
\alpha>0,
\end{equation}
where $\Gamma(x)$ denotes the Gamma function,
\begin{equation}\label{spg2}
\Gamma(x)
=
\int_{0}^{\infty}
t^{x-1}e^{-t}\,dt,
\qquad
x>0,
\end{equation} 
plays a central role in fractional analysis as a natural generalization of the exponential function \cite{parida2025mittag, abramowitz1965handbook, Gorenflo2020}. Depending on the value of the parameter $\alpha$, the ML function interpolates continuously between exponential-type and power-law behavior. This property makes it particularly suitable for constructing generalized hyperbolic structures and motivates its use in the present work as the basis for a super--hyperbolic deformation of catenoidal wormhole geometries.

\subsection{\label{subsec:level1.2}Novelty and Significance}

Motivated by the preceding discussion, the present work investigates a generalized deformation of the classical catenoidal wormhole geometry by introducing a \emph{super hyperbolic cosine function} (SHCF), denoted by $\mathfrak{S}C_{\alpha}(v)$, constructed from the ML function \cite{yang2021theory}. The function is defined as
\begin{equation}\label{1.2}
    \mathfrak{S}C_{\alpha} (v) = E_{2\alpha}\!\left(|v|^{2\alpha}\right) - 1 = \sum_{n=1}^{\infty} \frac{|v|^{2n\alpha}}{\Gamma(2n\alpha + 1)},
\end{equation}
where $E_{2\alpha}(z)$ is the ML function of Eq.~(\ref{spg1}) evaluated at order $2\alpha$, and $\Gamma(x)$ is the Gamma function; the argument is taken as $|v|^{2n\alpha}$ rather than $v^{2n\alpha}$ so that the series is real-valued and manifestly even for every real $v$, with no branch ambiguity. In the limit $\alpha\rightarrow1^{+}$, Eq.~(\ref{1.2}) approaches $\mathfrak{S}C_{\alpha}(v)\rightarrow\cosh(v)-1$; since the present construction additionally requires a non-degenerate local curvature term at the throat (Sec.~\ref{sec:level2}), the classical catenoidal embedding is recovered only as a boundary limit of the admissible parameter space, rather than at an interior value of $\alpha$. For $\alpha>1$, the Mittag-Leffler structure defines a higher-order deformation of the throat geometry, governed by the radial scale at which it becomes significant relative to the throat's local curvature, rather than by any non-integer behavior at $v=0$ itself.
Within the present framework, the non-degenerate curvature required at the throat is supplied by an explicit local term rather than by the Mittag-Leffler structure itself, so that regularity no longer depends on the value of $\alpha$. The parameter $\alpha$ instead controls two distinct higher-order aspects of the geometry: the radial scale at which the deformation departs from the purely local throat behavior, and the amplitude of the embedding's growth in the far field, whose asymptotic rate is fixed independently of $\alpha$. In this sense, the proposed construction extends the standard hyperbolic embedding into a continuously deformable family of generalized throat geometries, with the local and extended behavior of the throat now governed by distinct parameters.

The principal novelty of the present work lies in combining three interconnected ingredients within a single geometrical framework:
\begin{enumerate}
    \item a generalized super-hyperbolic embedding generated through the ML structure,
    \item a non-constant redshift function governing tidal effects,
    \item and a thin-shell stability analysis derived from the resulting spacetime geometry.
\end{enumerate}

Unlike conventional catenoidal wormhole constructions based on the classical functions $\cosh(v)$ and $\sinh(v)$, the present model separates the embedding into an explicit local curvature term, fixed by elementary regularity requirements, and a continuous deformation parameter $\alpha>1$ governing the embedding's behavior away from the throat. The corresponding spacetime metric therefore interpolates continuously between a near-throat geometry of guaranteed regularity and progressively more extended, Mittag-Leffler-dominated configurations at larger radial distance, approaching the classical catenoidal limit as $\alpha\rightarrow1^{+}$.

The present analysis is intended primarily as a geometrical and gravitational investigation of how generalized hyperbolic embeddings influence wormhole curvature, energy conditions, and stability properties within general relativity. In particular, the role of the deformation parameter $\alpha$ is studied through its influence on the onset radius of the deformation, the far-field growth of the embedding, and the resulting stress-energy and stability properties away from the throat, while the flare-out condition itself is shown to hold independently of $\alpha$. Although the ML deformation introduces features commonly associated with fractional or non-local geometric behavior, the present work does not assume a fundamentally fractional spacetime theory. Rather, the deformation is employed here as a generalized geometric embedding framework whose physical consequences are analyzed systematically within the classical Einstein field equations.

Accordingly, the significance of the model lies in providing a mathematically explicit extension of the catenoidal wormhole geometry in which the throat curvature is fixed by an independent, non-degenerate local parameter, while the extended profile and stability characteristics may be continuously tuned through the deformation parameter $\alpha$. This allows the investigation of how generalized embedding deformations modify the geometrical and physical properties of traversable wormholes beyond the standard MT construction.

\subsection{\label{subsec:level1.3}Structure of the Paper}

The paper is organized in a self-contained way, building the theory from geometric grounds up to physical implications. Section 2 establishes the mathematical preliminaries of the super--hyperbolic functions. This section lays down the analytic ground for the \emph{embedding function $\varrho (v)$}. Section 3 builds up the embedding and the wormhole metric. We start from the rotational embedding in Euclidean space and work out the induced metric, extracting the \emph{shape function} $b(r)$ and its derivatives. The appearance of the flare--out condition is shown explicitly. Section 4 introduces the non--constant \emph{redshift function} $\Phi(r)$ and incorporates it into the spacetime metric. We calculate its derivatives and then show how the ensuing field equations generalize the classical MT system but include tidal contributions through $\Phi^{\backprime} (r)$ and $\Phi^{\backprime \backprime} (r)$ (here $^\backprime$ indicates the derivative with respect to $r$ and through out the paper). Section 5 deduces the EFEs for the anisotropic matter distribution supporting the wormhole. The energy condition analysis is also performed in this section, followed by the evaluation of the null, weak, and strong energy conditions at and near the throat. In Section 6, we construct the thin--shell formalism and examine stability analysis using the \emph{Israel--Lanczos junction} conditions \cite{israel1966singular}, matching the interior super--hyperbolic wormhole to an exterior Schwarzschild solution.

\section{\label{sec:level2}Mathematical Preliminaries}
\subsection{\label{subsec:level2.1}Embedding of Catenoid and its Generalization}

The catenoid is recognized within the field of classical differential geometry as the inaugural example of a minimal surface characterized by revolution \cite{abbena2017modern, osserman2013survey}. This surface is constructed through the process of rotating a catenary curve about its axis of symmetry. In established mathematical terminology, the catenary \cite{choudhury2025catenoid} is expressed by the equation
\begin{equation}\label{2.1}
    r(z) = a \cosh \left(\frac{z}{a}\right),
\end{equation}
where the constant parameter $a>0$ determines the characteristic geometric scale of the surface and fixes the minimum neck radius of the corresponding catenoid. The associated surface of revolution may therefore be parametrized as
\begin{equation}\label{2.2}
(x,y,z)
=
\left(
a\cosh\left(\frac{v}{a}\right)\cos u,\,
a\cosh\left(\frac{v}{a}\right)\sin u,\,
av
\right),
\end{equation}
where
\[
u\in[0,2\pi),
\qquad
v\in\mathbb{R}.
\]
The induced metric on the catenoidal surface then takes the form
\begin{equation}\label{2.3}
ds_{\mathrm{cat}}^{2}
=
\cosh^{2}\left(\frac{v}{a}\right)
\left(
a^{2}du^{2}+dv^{2}
\right).
\end{equation}
The corresponding mean curvature vanishes identically,
\[
H=0,
\]
thereby confirming the catenoid as a minimal surface.

The classical catenoidal geometry is generated entirely by the integer-order hyperbolic function $\cosh(v)$, whose curvature structure is determined by ordinary differential behavior. In the present work, we investigate a generalized embedding obtained through a super--hyperbolic deformation governed by a Mittag--Leffler-type expansion. The objective is to introduce an additional geometric deformation, governed by a continuous parameter, that extends the embedding beyond the classical catenoidal profile while preserving its essential rotational structure; the local curvature required at the throat is secured independently, as shown below.

To this end, we replace the classical hyperbolic cosine by the super--hyperbolic cosine function (SHCF) $\mathfrak{S}C_{\alpha}(v)$ defined through the Mittag--Leffler-type series
\begin{equation}\label{2.4}
\mathfrak{S}C_{\alpha}(v)
=
E_{2\alpha}\!\left(|v|^{2\alpha}\right)-1
=
\sum_{n=1}^{\infty}
\frac{|v|^{2n\alpha}}{\Gamma(2n\alpha+1)},
\qquad
\alpha>1,
\end{equation}
where $E_{2\alpha}(z)$ is the ML function of Eq.~\eqref{spg1} evaluated at order $2\alpha$, and $\Gamma(x)$ denotes the Gamma function. The argument is taken as $|v|^{2n\alpha}$, rather than $v^{2n\alpha}$, so that the series is real-valued for every real $v$ without any branch ambiguity in the non-integer powers, and the sum begins at $n=1$ so that $\mathfrak{S}C_{\alpha}(0)=0$ identically. The parameter
\[
\alpha>1
\]
acts as a geometric deformation parameter controlling the higher-order scaling of the embedding away from the throat; the reason the admissible range excludes $\alpha\leq1$ is established in Sec.~\ref{subsec:lavel2.3}.
As $\alpha\rightarrow1^{+}$, the ML expansion approaches the ordinary hyperbolic cosine function shifted by its constant term \cite{yang2021theory},
\begin{equation}\label{2.5}
\lim_{\alpha\rightarrow1^{+}}
\mathfrak{S}C_{\alpha}(v)
=
\cosh(v)-1,
\end{equation}
so that the classical catenoidal profile is approached only as a boundary limit of the admissible parameter space, in conjunction with the local curvature construction of Sec.~\ref{subsec:lavel2.2}, rather than being attained at $\alpha=1$ itself.

The derivatives of $\mathfrak{S}C_{\alpha}(v)$ are obtained through term-by-term differentiation of the convergent series, written in terms of $|v|$ so as to remain real-valued for every real $v$:
\begin{equation}\label{2.6}
\left.
\begin{aligned}
\mathfrak{S}C_{\alpha}^{\prime}(v)
&=
\mathrm{sgn}(v)\sum_{n=1}^{\infty}
\frac{|v|^{2n\alpha-1}}{\Gamma(2n\alpha)},
\\[0.3cm]
\mathfrak{S}C_{\alpha}^{\prime\prime}(v)
&=
\sum_{n=1}^{\infty}
\frac{|v|^{2n\alpha-2}}{\Gamma(2n\alpha-1)},
\\[0.3cm]
\mathfrak{S}C_{\alpha}^{(m)}(v)
&=
\bigl[\mathrm{sgn}(v)\bigr]^{m}\sum_{n=1}^{\infty}
\frac{|v|^{2n\alpha-m}}
{\Gamma(2n\alpha-m+1)}.
\end{aligned}
\right\}
\end{equation}
Throughout this paper, the notation ${}^{\prime}$ denotes differentiation with respect to the embedding coordinate $v$, unless stated otherwise explicitly, and ${}^{(m)}$ denotes the $m$-th derivative with respect to $v$; the factor $[\mathrm{sgn}(v)]^{m}$ is understood as $\mathrm{sgn}(v)$ for odd $m$ and $1$ for even $m$, consistent with $\mathfrak{S}C_{\alpha}(v)$ being an even function of $v$.

At the throat location $v=0$, since $\alpha>1$ makes every exponent in Eqs.~\eqref{2.4} and \eqref{2.6} strictly positive, one obtains
\begin{equation}\label{2.7}
\mathfrak{S}C_{\alpha}(0)=0,
\qquad
\mathfrak{S}C_{\alpha}^{\prime}(0)=0,
\qquad
\mathfrak{S}C_{\alpha}^{\prime\prime}(0)=0.
\end{equation}
That is, for any $\alpha>1$, the Mittag--Leffler deformation alone has a degenerate stationary point at $v=0$: it contributes neither height, slope, nor curvature there. This is the reason a purely Mittag--Leffler embedding cannot by itself supply a genuine throat, and it motivates the explicit local curvature term introduced in Sec.~\ref{subsec:lavel2.2}.

The leading-order behavior of the second derivative near the throat is governed by the $n=1$ term of Eq.~\eqref{2.6},
\begin{equation}\label{2.8}
\mathfrak{S}C_{\alpha}^{\prime\prime}(v)
\sim
\frac{|v|^{2\alpha-2}}
{\Gamma(2\alpha-1)},
\qquad
v\rightarrow0,
\end{equation}
which vanishes as $v\rightarrow0$ for every $\alpha>1$, consistent with Eq.~\eqref{2.7}. Because this leading term grows away from the throat, it is not negligible at every radius; the radial scale at which it becomes significant relative to a given local throat curvature is established in Sec.~\ref{subsec:lavel2.3}.

Since every coefficient $1/\Gamma(2n\alpha-1)$ in Eq.~\eqref{2.6} is strictly positive for $\alpha>1$, it follows that $\mathfrak{S}C_{\alpha}^{\prime\prime}(v)>0$ for every $v\neq0$, vanishing only at the throat itself. The Mittag--Leffler contribution to the embedding curvature is therefore sign-definite throughout, rather than only in a restricted neighborhood of the throat; the corresponding statement for the full embedding $\varrho(v)$, once the local curvature term is included, is established in Sec.~\ref{subsec:lavel2.2}.

Unlike the classical case, in which the parameter enters only through the overall throat scale $a$, here $\alpha$ enters as an exponent and therefore reshapes the functional form of the deformation itself. Since $\mathfrak{S}C_{\alpha}(v)$ and its first two derivatives all vanish at $v=0$ for $\alpha>1$, this reshaping has no effect on the throat's local geometry; its influence is confined to how the embedding departs from that local geometry away from the throat, and to its growth at large $|v|$. Both effects are quantified explicitly in Sec.~\ref{subsec:lavel2.3}.

\subsection{\label{subsec:lavel2.2} Surface of Revolution}

We now construct the corresponding surface of revolution associated with the generalized super--hyperbolic embedding introduced previously. As established in Sec.~\ref{subsec:level2.1}, the function $\mathfrak{S}C_{\alpha}(v)$ alone has a degenerate stationary point at $v=0$ for every $\alpha>1$ (Eq.~\eqref{2.7}), so it cannot by itself define a genuine throat. We therefore first specify the minimal geometric requirements that any admissible throat embedding must satisfy, independently of the Mittag--Leffler structure.

A smooth radial embedding $\varrho:\mathbb{R}\rightarrow\mathbb{R}^{+}$ is called \emph{admissible} if it satisfies:
\begin{description}
    \item[Axiom 1] $\varrho(0)=\varrho_{0}>0$ (existence of a throat);
    \item[Axiom 2] $\varrho'(0)=0$ (stationary throat);
    \item[Axiom 3] $\varrho''(0)>0$ (strict local minimum);
    \item[Axiom 4] $\varrho\in C^{2}(\mathbb{R})$ (regularity sufficient for the curvature quantities used in this work);
    \item[Axiom 5] $\varrho$ is monotonically increasing for $v>0$ (monotonicity away from the throat);
    \item[Axiom 6] the induced curvature quantities of Sec.~\ref{sec:level5} remain finite for every finite $v$ (regular intrinsic geometry);
    \item[Axiom 7] $\varrho(v)$ admits a consistent exterior matching in the sense of Sec.~\ref{sec:level6} (asymptotically admissible growth).
\end{description}
Let $\mathcal{A}=\{\varrho(v):\varrho\text{ admissible}\}$ denote the space of admissible analytic embeddings.

By Taylor's theorem, any $\varrho\in C^{2}(\mathbb{R})$ satisfying Axiom 2 admits the local expansion $\varrho(v)=\varrho(0)+\tfrac{1}{2}\varrho''(0)v^{2}+\mathcal{O}(v^{3})$, so that Axiom 3 requires a genuinely non-vanishing quadratic term; this term is not an additional assumption but a necessary consequence of demanding a strict local minimum. Since Eq.~\eqref{2.7} shows that $\mathfrak{S}C_{\alpha}(v)$ contributes nothing at this order, the required quadratic term must be supplied independently. We accordingly decompose the embedding as
\begin{equation}\label{2.4a}
\varrho(v)=\varrho_{\mathrm{local}}(v)+\varrho_{\mathrm{global}}(v),
\qquad
\varrho_{\mathrm{local}}(v)=\varrho_{0}+\kappa v^{2},
\qquad
\varrho_{\mathrm{global}}(v)=a\,\mathfrak{S}C_{\alpha}(v),
\end{equation}
with $\kappa>0$ identified as $\kappa=\varrho''(0)/2$, so that
\begin{equation}\label{2.4.1}
\varrho(v)
=
\varrho_{0}+\kappa v^{2}+a\,\mathfrak{S}C_{\alpha}(v),
\qquad
z=2av.
\end{equation}
Here $\varrho_{0}=\varrho(0)>0$ is the throat radius, $\kappa>0$ is the local throat-curvature parameter, and $a>0$ retains its role as the geometric scaling parameter of the Mittag--Leffler deformation. Unless stated otherwise, differentiation with respect to the embedding coordinate $v$ will be denoted throughout this section by
\[
\varrho'(v)\equiv \frac{d\varrho}{dv}.
\]

The corresponding rotational embedding in $\mathbb{R}^{3}$ is defined parametrically by
\begin{equation}\label{2.8.1}
\left.
\begin{aligned}
   x &= \varrho(v)\cos u, \\
   y &= \varrho(v)\sin u, \\
   z &= 2av,
\end{aligned}
\right\}
\end{equation}
with coordinates
\[
(u,v)\in [0,2\pi)\times(-\infty,\infty).
\]
The parameter $u$ denotes the azimuthal angular coordinate associated with the rotational symmetry of the surface, while $v$ parametrizes the longitudinal embedding direction along the throat geometry.

The tangent vectors associated with the parametrization \eqref{2.8.1} are
\begin{equation}\label{2.8.2}
\mathbf{X}_{u}
=
\left(
-\varrho(v)\sin u,\,
\varrho(v)\cos u,\,
0
\right),
\end{equation}
and
\begin{equation}\label{2.8.3}
\mathbf{X}_{v}
=
\left(
\varrho'(v)\cos u,\,
\varrho'(v)\sin u,\,
2a
\right).
\end{equation}

Using the Euclidean metric in $\mathbb{R}^{3}$, the induced metric on the embedded surface is obtained from the first fundamental form,
\[
ds^{2}
=
\mathbf{X}_{u}\cdot \mathbf{X}_{u}\,du^{2}
+
2\,\mathbf{X}_{u}\cdot \mathbf{X}_{v}\,du\,dv
+
\mathbf{X}_{v}\cdot \mathbf{X}_{v}\,dv^{2}.
\]
Direct computation gives
\begin{equation}\label{2.8.4}
\mathbf{X}_{u}\cdot \mathbf{X}_{u}
=
\varrho(v)^{2},
\end{equation}
\begin{equation}\label{2.8.5}
\mathbf{X}_{u}\cdot \mathbf{X}_{v}
=
0,
\end{equation}
and
\begin{equation}\label{2.8.6}
\mathbf{X}_{v}\cdot \mathbf{X}_{v}
=
\bigl(\varrho'(v)\bigr)^{2}
+
4a^{2}.
\end{equation}

Consequently, the induced surface metric becomes
\begin{equation}\label{2.9}
    ds_{shc}^{2}
    =
    \varrho(v)^{2}\,du^{2}
    +
    \left[
    \bigl(\varrho'(v)\bigr)^{2}
    +
    4a^{2}
    \right]dv^{2}.
\end{equation}

Equation \eqref{2.9} represents the intrinsic geometry of the generalized super--hyperbolic surface generated by the rotational embedding \eqref{2.8.1}. In the boundary limit $\alpha\rightarrow1^{+}$, together with a vanishing relative contribution from the local curvature term $\kappa v^{2}$, the super--hyperbolic function approaches $\cosh(v)-1$ and the geometry approaches the classical catenoidal embedding.

To visualize the embedding geometry in cylindrical coordinates $(r,\theta,z)$ of the ambient Euclidean space, we identify
\begin{equation}\label{2.9a}
    r=\varrho(v),
    \qquad
    \theta=u,
    \qquad
    z=2av.
\end{equation}
The Euclidean line element of the surrounding three--space is therefore
\begin{equation}\label{2.10}
    dl^{2}
    =
    dr^{2}
    +
    r^{2}d\theta^{2}
    +
    dz^{2}.
\end{equation}

Using
\begin{equation}\label{2.10a}
    dr=\varrho'(v)\,dv,
\end{equation}
together with
\begin{equation}\label{2.10b}
    dz=2a\,dv,
\end{equation}
the line element \eqref{2.10} reduces identically to the induced surface metric \eqref{2.9}. Thus, the generalized super--hyperbolic geometry may be interpreted as a rotationally symmetric embedded surface whose intrinsic metric is completely determined by the embedding profile $\varrho(v)$.

\subsection{\label{subsec:lavel2.3}Analyticity and Regularity of the Embedding Function}

The embedding profile $\varrho(v)$ defined in Eq.~\eqref{2.4.1} has derivative
\begin{equation}\label{2.11}
\frac{d\varrho}{dv}
=
2\kappa v+a\,\mathfrak{S}C_{\alpha}^{\prime}(v)
=
2\kappa v+a\,\mathrm{sgn}(v)\sum_{n=1}^{\infty}
\frac{|v|^{2n\alpha-1}}{\Gamma(2n\alpha)},
\end{equation}
and second derivative
\begin{equation}\label{2.11a}
\frac{d^{2}\varrho}{dv^{2}}
=
2\kappa+a\,\mathfrak{S}C_{\alpha}^{\prime\prime}(v),
\end{equation}
where $\kappa>0$ is the local throat-curvature parameter of Eq.~\eqref{2.4a}, $a>0$ is the geometric scaling parameter, and $\alpha>1$ is the deformation parameter. Formally, the inverse relation associated with Eq.~\eqref{2.11} allows one to determine $v=v(\varrho)$ locally within regions where $d\varrho/dv \neq 0$.

\begin{figure}
\centering
\begin{subfigure}{0.48\linewidth}
\centering
\begin{tikzpicture}
\begin{axis}[
    width=\linewidth, height=7.2cm,
    view={-35}{18},
    hide axis,
    colormap={wormblue}{rgb255(0cm)=(222,235,247) rgb255(1cm)=(158,202,225)
                        rgb255(2cm)=(66,146,198) rgb255(3cm)=(8,69,148)},
    faceted color=black!40,
]
\addplot3[surf, shader=faceted interp, line width=0.15pt, mesh/rows=20] table {
7.871 0.0 -5.4
6.531 0.0 -5.0
5.407 0.0 -4.6
4.467 0.0 -4.2
3.682 0.0 -3.8
3.032 0.0 -3.4
2.497 0.0 -3.0
2.063 0.0 -2.6
1.717 0.0 -2.2
1.451 0.0 -1.8
1.254 0.0 -1.4
1.12 0.0 -1.0
1.039 0.0 -0.6
1.004 0.0 -0.2
1.004 0.0 0.2
1.039 0.0 0.6
1.12 0.0 1.0
1.254 0.0 1.4
1.451 0.0 1.8
1.717 0.0 2.2
2.063 0.0 2.6
2.497 0.0 3.0
3.032 0.0 3.4
3.682 0.0 3.8
4.467 0.0 4.2
5.407 0.0 4.6
6.531 0.0 5.0
7.871 0.0 5.4

7.444 2.556 -5.4
6.177 2.121 -5.0
5.114 1.756 -4.6
4.225 1.45 -4.2
3.483 1.196 -3.8
2.867 0.9844 -3.4
2.361 0.8106 -3.0
1.951 0.6697 -2.6
1.624 0.5577 -2.2
1.372 0.4711 -1.8
1.186 0.4073 -1.4
1.059 0.3635 -1.0
0.9825 0.3373 -0.6
0.9493 0.3259 -0.2
0.9493 0.3259 0.2
0.9825 0.3373 0.6
1.059 0.3635 1.0
1.186 0.4073 1.4
1.372 0.4711 1.8
1.624 0.5577 2.2
1.951 0.6697 2.6
2.361 0.8106 3.0
2.867 0.9844 3.4
3.483 1.196 3.8
4.225 1.45 4.2
5.114 1.756 4.6
6.177 2.121 5.0
7.444 2.556 5.4

6.211 4.834 -5.4
5.154 4.011 -5.0
4.267 3.321 -4.6
3.525 2.744 -4.2
2.906 2.262 -3.8
2.392 1.862 -3.4
1.97 1.533 -3.0
1.628 1.267 -2.6
1.355 1.055 -2.2
1.145 0.8912 -1.8
0.9899 0.7704 -1.4
0.8835 0.6876 -1.0
0.8197 0.638 -0.6
0.792 0.6165 -0.2
0.792 0.6165 0.2
0.8197 0.638 0.6
0.8835 0.6876 1.0
0.9899 0.7704 1.4
1.145 0.8912 1.8
1.355 1.055 2.2
1.628 1.267 2.6
1.97 1.533 3.0
2.392 1.862 3.4
2.906 2.262 3.8
3.525 2.744 4.2
4.267 3.321 4.6
5.154 4.011 5.0
6.211 4.834 5.4

4.305 6.589 -5.4
3.572 5.467 -5.0
2.958 4.527 -4.6
2.443 3.74 -4.2
2.014 3.083 -3.8
1.658 2.538 -3.4
1.365 2.09 -3.0
1.128 1.727 -2.6
0.9394 1.438 -2.2
0.7936 1.215 -1.8
0.6861 1.05 -1.4
0.6123 0.9372 -1.0
0.5682 0.8696 -0.6
0.549 0.8402 -0.2
0.549 0.8402 0.2
0.5682 0.8696 0.6
0.6123 0.9372 1.0
0.6861 1.05 1.4
0.7936 1.215 1.8
0.9394 1.438 2.2
1.128 1.727 2.6
1.365 2.09 3.0
1.658 2.538 3.4
2.014 3.083 3.8
2.443 3.74 4.2
2.958 4.527 4.6
3.572 5.467 5.0
4.305 6.589 5.4

1.932 7.63 -5.4
1.603 6.331 -5.0
1.327 5.242 -4.6
1.097 4.33 -4.2
0.904 3.57 -3.8
0.7442 2.939 -3.4
0.6129 2.42 -3.0
0.5063 1.999 -2.6
0.4216 1.665 -2.2
0.3562 1.407 -1.8
0.3079 1.216 -1.4
0.2748 1.085 -1.0
0.255 1.007 -0.6
0.2464 0.973 -0.2
0.2464 0.973 0.2
0.255 1.007 0.6
0.2748 1.085 1.0
0.3079 1.216 1.4
0.3562 1.407 1.8
0.4216 1.665 2.2
0.5063 1.999 2.6
0.6129 2.42 3.0
0.7442 2.939 3.4
0.904 3.57 3.8
1.097 4.33 4.2
1.327 5.242 4.6
1.603 6.331 5.0
1.932 7.63 5.4

-0.6499 7.844 -5.4
-0.5393 6.509 -5.0
-0.4465 5.389 -4.6
-0.3689 4.452 -4.2
-0.3041 3.67 -3.8
-0.2503 3.021 -3.4
-0.2062 2.488 -3.0
-0.1703 2.056 -2.6
-0.1418 1.712 -2.2
-0.1198 1.446 -1.8
-0.1036 1.25 -1.4
-0.09245 1.116 -1.0
-0.08578 1.035 -0.6
-0.08288 1.0 -0.2
-0.08288 1.0 0.2
-0.08578 1.035 0.6
-0.09245 1.116 1.0
-0.1036 1.25 1.4
-0.1198 1.446 1.8
-0.1418 1.712 2.2
-0.1703 2.056 2.6
-0.2062 2.488 3.0
-0.2503 3.021 3.4
-0.3041 3.67 3.8
-0.3689 4.452 4.2
-0.4465 5.389 4.6
-0.5393 6.509 5.0
-0.6499 7.844 5.4

-3.162 7.208 -5.4
-2.623 5.981 -5.0
-2.172 4.952 -4.6
-1.794 4.091 -4.2
-1.479 3.372 -3.8
-1.218 2.776 -3.4
-1.003 2.286 -3.0
-0.8285 1.889 -2.6
-0.6899 1.573 -2.2
-0.5829 1.329 -1.8
-0.5039 1.149 -1.4
-0.4497 1.025 -1.0
-0.4173 0.9513 -0.6
-0.4032 0.9191 -0.2
-0.4032 0.9191 0.2
-0.4173 0.9513 0.6
-0.4497 1.025 1.0
-0.5039 1.149 1.4
-0.5829 1.329 1.8
-0.6899 1.573 2.2
-0.8285 1.889 2.6
-1.003 2.286 3.0
-1.218 2.776 3.4
-1.479 3.372 3.8
-1.794 4.091 4.2
-2.172 4.952 4.6
-2.623 5.981 5.0
-3.162 7.208 5.4

-5.331 5.791 -5.4
-4.423 4.805 -5.0
-3.662 3.978 -4.6
-3.025 3.286 -4.2
-2.494 2.709 -3.8
-2.053 2.23 -3.4
-1.691 1.837 -3.0
-1.397 1.517 -2.6
-1.163 1.264 -2.2
-0.9827 1.068 -1.8
-0.8496 0.9229 -1.4
-0.7582 0.8237 -1.0
-0.7035 0.7642 -0.6
-0.6798 0.7384 -0.2
-0.6798 0.7384 0.2
-0.7035 0.7642 0.6
-0.7582 0.8237 1.0
-0.8496 0.9229 1.4
-0.9827 1.068 1.8
-1.163 1.264 2.2
-1.397 1.517 2.6
-1.691 1.837 3.0
-2.053 2.23 3.4
-2.494 2.709 3.8
-3.025 3.286 4.2
-3.662 3.978 4.6
-4.423 4.805 5.0
-5.331 5.791 5.4

-6.922 3.746 -5.4
-5.744 3.108 -5.0
-4.756 2.574 -4.6
-3.929 2.126 -4.2
-3.239 1.753 -3.8
-2.666 1.443 -3.4
-2.196 1.188 -3.0
-1.814 0.9817 -2.6
-1.51 0.8174 -2.2
-1.276 0.6906 -1.8
-1.103 0.597 -1.4
-0.9846 0.5328 -1.0
-0.9136 0.4944 -0.6
-0.8827 0.4777 -0.2
-0.8827 0.4777 0.2
-0.9136 0.4944 0.6
-0.9846 0.5328 1.0
-1.103 0.597 1.4
-1.276 0.6906 1.8
-1.51 0.8174 2.2
-1.814 0.9817 2.6
-2.196 1.188 3.0
-2.666 1.443 3.4
-3.239 1.753 3.8
-3.929 2.126 4.2
-4.756 2.574 4.6
-5.744 3.108 5.0
-6.922 3.746 5.4

-7.763 1.295 -5.4
-6.442 1.075 -5.0
-5.334 0.89 -4.6
-4.406 0.7352 -4.2
-3.632 0.6061 -3.8
-2.99 0.499 -3.4
-2.463 0.4109 -3.0
-2.034 0.3395 -2.6
-1.694 0.2827 -2.2
-1.431 0.2388 -1.8
-1.237 0.2065 -1.4
-1.104 0.1843 -1.0
-1.025 0.171 -0.6
-0.99 0.1652 -0.2
-0.99 0.1652 0.2
-1.025 0.171 0.6
-1.104 0.1843 1.0
-1.237 0.2065 1.4
-1.431 0.2388 1.8
-1.694 0.2827 2.2
-2.034 0.3395 2.6
-2.463 0.4109 3.0
-2.99 0.499 3.4
-3.632 0.6061 3.8
-4.406 0.7352 4.2
-5.334 0.89 4.6
-6.442 1.075 5.0
-7.763 1.295 5.4

-7.763 -1.295 -5.4
-6.442 -1.075 -5.0
-5.334 -0.89 -4.6
-4.406 -0.7352 -4.2
-3.632 -0.6061 -3.8
-2.99 -0.499 -3.4
-2.463 -0.4109 -3.0
-2.034 -0.3395 -2.6
-1.694 -0.2827 -2.2
-1.431 -0.2388 -1.8
-1.237 -0.2065 -1.4
-1.104 -0.1843 -1.0
-1.025 -0.171 -0.6
-0.99 -0.1652 -0.2
-0.99 -0.1652 0.2
-1.025 -0.171 0.6
-1.104 -0.1843 1.0
-1.237 -0.2065 1.4
-1.431 -0.2388 1.8
-1.694 -0.2827 2.2
-2.034 -0.3395 2.6
-2.463 -0.4109 3.0
-2.99 -0.499 3.4
-3.632 -0.6061 3.8
-4.406 -0.7352 4.2
-5.334 -0.89 4.6
-6.442 -1.075 5.0
-7.763 -1.295 5.4

-6.922 -3.746 -5.4
-5.744 -3.108 -5.0
-4.756 -2.574 -4.6
-3.929 -2.126 -4.2
-3.239 -1.753 -3.8
-2.666 -1.443 -3.4
-2.196 -1.188 -3.0
-1.814 -0.9817 -2.6
-1.51 -0.8174 -2.2
-1.276 -0.6906 -1.8
-1.103 -0.597 -1.4
-0.9846 -0.5328 -1.0
-0.9136 -0.4944 -0.6
-0.8827 -0.4777 -0.2
-0.8827 -0.4777 0.2
-0.9136 -0.4944 0.6
-0.9846 -0.5328 1.0
-1.103 -0.597 1.4
-1.276 -0.6906 1.8
-1.51 -0.8174 2.2
-1.814 -0.9817 2.6
-2.196 -1.188 3.0
-2.666 -1.443 3.4
-3.239 -1.753 3.8
-3.929 -2.126 4.2
-4.756 -2.574 4.6
-5.744 -3.108 5.0
-6.922 -3.746 5.4

-5.331 -5.791 -5.4
-4.423 -4.805 -5.0
-3.662 -3.978 -4.6
-3.025 -3.286 -4.2
-2.494 -2.709 -3.8
-2.053 -2.23 -3.4
-1.691 -1.837 -3.0
-1.397 -1.517 -2.6
-1.163 -1.264 -2.2
-0.9827 -1.068 -1.8
-0.8496 -0.9229 -1.4
-0.7582 -0.8237 -1.0
-0.7035 -0.7642 -0.6
-0.6798 -0.7384 -0.2
-0.6798 -0.7384 0.2
-0.7035 -0.7642 0.6
-0.7582 -0.8237 1.0
-0.8496 -0.9229 1.4
-0.9827 -1.068 1.8
-1.163 -1.264 2.2
-1.397 -1.517 2.6
-1.691 -1.837 3.0
-2.053 -2.23 3.4
-2.494 -2.709 3.8
-3.025 -3.286 4.2
-3.662 -3.978 4.6
-4.423 -4.805 5.0
-5.331 -5.791 5.4

-3.162 -7.208 -5.4
-2.623 -5.981 -5.0
-2.172 -4.952 -4.6
-1.794 -4.091 -4.2
-1.479 -3.372 -3.8
-1.218 -2.776 -3.4
-1.003 -2.286 -3.0
-0.8285 -1.889 -2.6
-0.6899 -1.573 -2.2
-0.5829 -1.329 -1.8
-0.5039 -1.149 -1.4
-0.4497 -1.025 -1.0
-0.4173 -0.9513 -0.6
-0.4032 -0.9191 -0.2
-0.4032 -0.9191 0.2
-0.4173 -0.9513 0.6
-0.4497 -1.025 1.0
-0.5039 -1.149 1.4
-0.5829 -1.329 1.8
-0.6899 -1.573 2.2
-0.8285 -1.889 2.6
-1.003 -2.286 3.0
-1.218 -2.776 3.4
-1.479 -3.372 3.8
-1.794 -4.091 4.2
-2.172 -4.952 4.6
-2.623 -5.981 5.0
-3.162 -7.208 5.4

-0.6499 -7.844 -5.4
-0.5393 -6.509 -5.0
-0.4465 -5.389 -4.6
-0.3689 -4.452 -4.2
-0.3041 -3.67 -3.8
-0.2503 -3.021 -3.4
-0.2062 -2.488 -3.0
-0.1703 -2.056 -2.6
-0.1418 -1.712 -2.2
-0.1198 -1.446 -1.8
-0.1036 -1.25 -1.4
-0.09245 -1.116 -1.0
-0.08578 -1.035 -0.6
-0.08288 -1.0 -0.2
-0.08288 -1.0 0.2
-0.08578 -1.035 0.6
-0.09245 -1.116 1.0
-0.1036 -1.25 1.4
-0.1198 -1.446 1.8
-0.1418 -1.712 2.2
-0.1703 -2.056 2.6
-0.2062 -2.488 3.0
-0.2503 -3.021 3.4
-0.3041 -3.67 3.8
-0.3689 -4.452 4.2
-0.4465 -5.389 4.6
-0.5393 -6.509 5.0
-0.6499 -7.844 5.4

1.932 -7.63 -5.4
1.603 -6.331 -5.0
1.327 -5.242 -4.6
1.097 -4.33 -4.2
0.904 -3.57 -3.8
0.7442 -2.939 -3.4
0.6129 -2.42 -3.0
0.5063 -1.999 -2.6
0.4216 -1.665 -2.2
0.3562 -1.407 -1.8
0.3079 -1.216 -1.4
0.2748 -1.085 -1.0
0.255 -1.007 -0.6
0.2464 -0.973 -0.2
0.2464 -0.973 0.2
0.255 -1.007 0.6
0.2748 -1.085 1.0
0.3079 -1.216 1.4
0.3562 -1.407 1.8
0.4216 -1.665 2.2
0.5063 -1.999 2.6
0.6129 -2.42 3.0
0.7442 -2.939 3.4
0.904 -3.57 3.8
1.097 -4.33 4.2
1.327 -5.242 4.6
1.603 -6.331 5.0
1.932 -7.63 5.4

4.305 -6.589 -5.4
3.572 -5.467 -5.0
2.958 -4.527 -4.6
2.443 -3.74 -4.2
2.014 -3.083 -3.8
1.658 -2.538 -3.4
1.365 -2.09 -3.0
1.128 -1.727 -2.6
0.9394 -1.438 -2.2
0.7936 -1.215 -1.8
0.6861 -1.05 -1.4
0.6123 -0.9372 -1.0
0.5682 -0.8696 -0.6
0.549 -0.8402 -0.2
0.549 -0.8402 0.2
0.5682 -0.8696 0.6
0.6123 -0.9372 1.0
0.6861 -1.05 1.4
0.7936 -1.215 1.8
0.9394 -1.438 2.2
1.128 -1.727 2.6
1.365 -2.09 3.0
1.658 -2.538 3.4
2.014 -3.083 3.8
2.443 -3.74 4.2
2.958 -4.527 4.6
3.572 -5.467 5.0
4.305 -6.589 5.4

6.211 -4.834 -5.4
5.154 -4.011 -5.0
4.267 -3.321 -4.6
3.525 -2.744 -4.2
2.906 -2.262 -3.8
2.392 -1.862 -3.4
1.97 -1.533 -3.0
1.628 -1.267 -2.6
1.355 -1.055 -2.2
1.145 -0.8912 -1.8
0.9899 -0.7704 -1.4
0.8835 -0.6876 -1.0
0.8197 -0.638 -0.6
0.792 -0.6165 -0.2
0.792 -0.6165 0.2
0.8197 -0.638 0.6
0.8835 -0.6876 1.0
0.9899 -0.7704 1.4
1.145 -0.8912 1.8
1.355 -1.055 2.2
1.628 -1.267 2.6
1.97 -1.533 3.0
2.392 -1.862 3.4
2.906 -2.262 3.8
3.525 -2.744 4.2
4.267 -3.321 4.6
5.154 -4.011 5.0
6.211 -4.834 5.4

7.444 -2.556 -5.4
6.177 -2.121 -5.0
5.114 -1.756 -4.6
4.225 -1.45 -4.2
3.483 -1.196 -3.8
2.867 -0.9844 -3.4
2.361 -0.8106 -3.0
1.951 -0.6697 -2.6
1.624 -0.5577 -2.2
1.372 -0.4711 -1.8
1.186 -0.4073 -1.4
1.059 -0.3635 -1.0
0.9825 -0.3373 -0.6
0.9493 -0.3259 -0.2
0.9493 -0.3259 0.2
0.9825 -0.3373 0.6
1.059 -0.3635 1.0
1.186 -0.4073 1.4
1.372 -0.4711 1.8
1.624 -0.5577 2.2
1.951 -0.6697 2.6
2.361 -0.8106 3.0
2.867 -0.9844 3.4
3.483 -1.196 3.8
4.225 -1.45 4.2
5.114 -1.756 4.6
6.177 -2.121 5.0
7.444 -2.556 5.4

7.871 -1.033e-20 -5.4
6.531 -8.574e-21 -5.0
5.407 -7.099e-21 -4.6
4.467 -5.864e-21 -4.2
3.682 -4.834e-21 -3.8
3.032 -3.98e-21 -3.4
2.497 -3.277e-21 -3.0
2.063 -2.708e-21 -2.6
1.717 -2.255e-21 -2.2
1.451 -1.905e-21 -1.8
1.254 -1.647e-21 -1.4
1.12 -1.47e-21 -1.0
1.039 -1.364e-21 -0.6
1.004 -1.318e-21 -0.2
1.004 -1.318e-21 0.2
1.039 -1.364e-21 0.6
1.12 -1.47e-21 1.0
1.254 -1.647e-21 1.4
1.451 -1.905e-21 1.8
1.717 -2.255e-21 2.2
2.063 -2.708e-21 2.6
2.497 -3.277e-21 3.0
3.032 -3.98e-21 3.4
3.682 -4.834e-21 3.8
4.467 -5.864e-21 4.2
5.407 -7.099e-21 4.6
6.531 -8.574e-21 5.0
7.871 -1.033e-20 5.4
};
\end{axis}
\end{tikzpicture}
\caption{$\alpha=1.3$}
\end{subfigure}
\hfill
\begin{subfigure}{0.48\linewidth}
\centering
\begin{tikzpicture}
\begin{axis}[
    width=\linewidth, height=7.2cm,
    view={-35}{18},
    hide axis,
    colormap={wormblue}{rgb255(0cm)=(222,235,247) rgb255(1cm)=(158,202,225)
                        rgb255(2cm)=(66,146,198) rgb255(3cm)=(8,69,148)},
    faceted color=black!40,
]
\addplot3[surf, shader=faceted interp, line width=0.15pt, mesh/rows=20] table {
4.987 0.0 -5.4
4.152 0.0 -5.0
3.469 0.0 -4.6
2.913 0.0 -4.2
2.462 0.0 -3.8
2.099 0.0 -3.4
1.809 0.0 -3.0
1.578 0.0 -2.6
1.397 0.0 -2.2
1.257 0.0 -1.8
1.152 0.0 -1.4
1.076 0.0 -1.0
1.027 0.0 -0.6
1.003 0.0 -0.2
1.003 0.0 0.2
1.027 0.0 0.6
1.076 0.0 1.0
1.152 0.0 1.4
1.257 0.0 1.8
1.397 0.0 2.2
1.578 0.0 2.6
1.809 0.0 3.0
2.099 0.0 3.4
2.462 0.0 3.8
2.913 0.0 4.2
3.469 0.0 4.6
4.152 0.0 5.0
4.987 0.0 5.4

4.717 1.619 -5.4
3.927 1.348 -5.0
3.281 1.126 -4.6
2.755 0.9457 -4.2
2.329 0.7994 -3.8
1.985 0.6815 -3.4
1.711 0.5873 -3.0
1.493 0.5124 -2.6
1.321 0.4536 -2.2
1.189 0.4082 -1.8
1.089 0.3739 -1.4
1.018 0.3494 -1.0
0.9715 0.3335 -0.6
0.9487 0.3257 -0.2
0.9487 0.3257 0.2
0.9715 0.3335 0.6
1.018 0.3494 1.0
1.089 0.3739 1.4
1.189 0.4082 1.8
1.321 0.4536 2.2
1.493 0.5124 2.6
1.711 0.5873 3.0
1.985 0.6815 3.4
2.329 0.7994 3.8
2.755 0.9457 4.2
3.281 1.126 4.6
3.927 1.348 5.0
4.717 1.619 5.4

3.935 3.063 -5.4
3.277 2.55 -5.0
2.738 2.131 -4.6
2.299 1.789 -4.2
1.943 1.512 -3.8
1.656 1.289 -3.4
1.427 1.111 -3.0
1.245 0.9693 -2.6
1.103 0.8581 -2.2
0.992 0.7721 -1.8
0.9088 0.7074 -1.4
0.8492 0.6609 -1.0
0.8105 0.6309 -0.6
0.7915 0.6161 -0.2
0.7915 0.6161 0.2
0.8105 0.6309 0.6
0.8492 0.6609 1.0
0.9088 0.7074 1.4
0.992 0.7721 1.8
1.103 0.8581 2.2
1.245 0.9693 2.6
1.427 1.111 3.0
1.656 1.289 3.4
1.943 1.512 3.8
2.299 1.789 4.2
2.738 2.131 4.6
3.277 2.55 5.0
3.935 3.063 5.4

2.728 4.175 -5.4
2.271 3.476 -5.0
1.897 2.904 -4.6
1.593 2.438 -4.2
1.347 2.061 -3.8
1.148 1.757 -3.4
0.9892 1.514 -3.0
0.8632 1.321 -2.6
0.7641 1.17 -2.2
0.6876 1.052 -1.8
0.6299 0.9641 -1.4
0.5886 0.9008 -1.0
0.5618 0.8599 -0.6
0.5486 0.8397 -0.2
0.5486 0.8397 0.2
0.5618 0.8599 0.6
0.5886 0.9008 1.0
0.6299 0.9641 1.4
0.6876 1.052 1.8
0.7641 1.17 2.2
0.8632 1.321 2.6
0.9892 1.514 3.0
1.148 1.757 3.4
1.347 2.061 3.8
1.593 2.438 4.2
1.897 2.904 4.6
2.271 3.476 5.0
2.728 4.175 5.4

1.224 4.834 -5.4
1.019 4.025 -5.0
0.8516 3.363 -4.6
0.715 2.824 -4.2
0.6044 2.387 -3.8
0.5153 2.035 -3.4
0.444 1.753 -3.0
0.3874 1.53 -2.6
0.343 1.354 -2.2
0.3086 1.219 -1.8
0.2827 1.116 -1.4
0.2642 1.043 -1.0
0.2521 0.9957 -0.6
0.2462 0.9723 -0.2
0.2462 0.9723 0.2
0.2521 0.9957 0.6
0.2642 1.043 1.0
0.2827 1.116 1.4
0.3086 1.219 1.8
0.343 1.354 2.2
0.3874 1.53 2.6
0.444 1.753 3.0
0.5153 2.035 3.4
0.6044 2.387 3.8
0.715 2.824 4.2
0.8516 3.363 4.6
1.019 4.025 5.0
1.224 4.834 5.4

-0.4118 4.97 -5.4
-0.3429 4.138 -5.0
-0.2865 3.457 -4.6
-0.2405 2.903 -4.2
-0.2033 2.454 -3.8
-0.1733 2.092 -3.4
-0.1494 1.802 -3.0
-0.1303 1.573 -2.6
-0.1154 1.392 -2.2
-0.1038 1.253 -1.8
-0.0951 1.148 -1.4
-0.08886 1.072 -1.0
-0.08482 1.024 -0.6
-0.08283 0.9996 -0.2
-0.08283 0.9996 0.2
-0.08482 1.024 0.6
-0.08886 1.072 1.0
-0.0951 1.148 1.4
-0.1038 1.253 1.8
-0.1154 1.392 2.2
-0.1303 1.573 2.6
-0.1494 1.802 3.0
-0.1733 2.092 3.4
-0.2033 2.454 3.8
-0.2405 2.903 4.2
-0.2865 3.457 4.6
-0.3429 4.138 5.0
-0.4118 4.97 5.4

-2.003 4.567 -5.4
-1.668 3.803 -5.0
-1.394 3.177 -4.6
-1.17 2.667 -4.2
-0.989 2.255 -3.8
-0.8432 1.922 -3.4
-0.7265 1.656 -3.0
-0.6339 1.445 -2.6
-0.5612 1.279 -2.2
-0.505 1.151 -1.8
-0.4626 1.055 -1.4
-0.4322 0.9854 -1.0
-0.4126 0.9406 -0.6
-0.4029 0.9185 -0.2
-0.4029 0.9185 0.2
-0.4126 0.9406 0.6
-0.4322 0.9854 1.0
-0.4626 1.055 1.4
-0.505 1.151 1.8
-0.5612 1.279 2.2
-0.6339 1.445 2.6
-0.7265 1.656 3.0
-0.8432 1.922 3.4
-0.989 2.255 3.8
-1.17 2.667 4.2
-1.394 3.177 4.6
-1.668 3.803 5.0
-2.003 4.567 5.4

-3.378 3.669 -5.4
-2.812 3.055 -5.0
-2.35 2.552 -4.6
-1.973 2.143 -4.2
-1.667 1.811 -3.8
-1.422 1.544 -3.4
-1.225 1.331 -3.0
-1.069 1.161 -2.6
-0.9462 1.028 -2.2
-0.8514 0.9249 -1.8
-0.78 0.8473 -1.4
-0.7288 0.7917 -1.0
-0.6956 0.7557 -0.6
-0.6793 0.7379 -0.2
-0.6793 0.7379 0.2
-0.6956 0.7557 0.6
-0.7288 0.7917 1.0
-0.78 0.8473 1.4
-0.8514 0.9249 1.8
-0.9462 1.028 2.2
-1.069 1.161 2.6
-1.225 1.331 3.0
-1.422 1.544 3.4
-1.667 1.811 3.8
-1.973 2.143 4.2
-2.35 2.552 4.6
-2.812 3.055 5.0
-3.378 3.669 5.4

-4.386 2.373 -5.4
-3.652 1.976 -5.0
-3.051 1.651 -4.6
-2.562 1.386 -4.2
-2.165 1.172 -3.8
-1.846 0.999 -3.4
-1.591 0.8608 -3.0
-1.388 0.7511 -2.6
-1.229 0.665 -2.2
-1.106 0.5983 -1.8
-1.013 0.5481 -1.4
-0.9464 0.5121 -1.0
-0.9033 0.4889 -0.6
-0.8821 0.4774 -0.2
-0.8821 0.4774 0.2
-0.9033 0.4889 0.6
-0.9464 0.5121 1.0
-1.013 0.5481 1.4
-1.106 0.5983 1.8
-1.229 0.665 2.2
-1.388 0.7511 2.6
-1.591 0.8608 3.0
-1.846 0.999 3.4
-2.165 1.172 3.8
-2.562 1.386 4.2
-3.051 1.651 4.6
-3.652 1.976 5.0
-4.386 2.373 5.4

-4.919 0.8208 -5.4
-4.096 0.6835 -5.0
-3.422 0.571 -4.6
-2.873 0.4794 -4.2
-2.428 0.4052 -3.8
-2.07 0.3455 -3.4
-1.784 0.2977 -3.0
-1.557 0.2598 -2.6
-1.378 0.23 -2.2
-1.24 0.2069 -1.8
-1.136 0.1896 -1.4
-1.061 0.1771 -1.0
-1.013 0.1691 -0.6
-0.9893 0.1651 -0.2
-0.9893 0.1651 0.2
-1.013 0.1691 0.6
-1.061 0.1771 1.0
-1.136 0.1896 1.4
-1.24 0.2069 1.8
-1.378 0.23 2.2
-1.557 0.2598 2.6
-1.784 0.2977 3.0
-2.07 0.3455 3.4
-2.428 0.4052 3.8
-2.873 0.4794 4.2
-3.422 0.571 4.6
-4.096 0.6835 5.0
-4.919 0.8208 5.4

-4.919 -0.8208 -5.4
-4.096 -0.6835 -5.0
-3.422 -0.571 -4.6
-2.873 -0.4794 -4.2
-2.428 -0.4052 -3.8
-2.07 -0.3455 -3.4
-1.784 -0.2977 -3.0
-1.557 -0.2598 -2.6
-1.378 -0.23 -2.2
-1.24 -0.2069 -1.8
-1.136 -0.1896 -1.4
-1.061 -0.1771 -1.0
-1.013 -0.1691 -0.6
-0.9893 -0.1651 -0.2
-0.9893 -0.1651 0.2
-1.013 -0.1691 0.6
-1.061 -0.1771 1.0
-1.136 -0.1896 1.4
-1.24 -0.2069 1.8
-1.378 -0.23 2.2
-1.557 -0.2598 2.6
-1.784 -0.2977 3.0
-2.07 -0.3455 3.4
-2.428 -0.4052 3.8
-2.873 -0.4794 4.2
-3.422 -0.571 4.6
-4.096 -0.6835 5.0
-4.919 -0.8208 5.4

-4.386 -2.373 -5.4
-3.652 -1.976 -5.0
-3.051 -1.651 -4.6
-2.562 -1.386 -4.2
-2.165 -1.172 -3.8
-1.846 -0.999 -3.4
-1.591 -0.8608 -3.0
-1.388 -0.7511 -2.6
-1.229 -0.665 -2.2
-1.106 -0.5983 -1.8
-1.013 -0.5481 -1.4
-0.9464 -0.5121 -1.0
-0.9033 -0.4889 -0.6
-0.8821 -0.4774 -0.2
-0.8821 -0.4774 0.2
-0.9033 -0.4889 0.6
-0.9464 -0.5121 1.0
-1.013 -0.5481 1.4
-1.106 -0.5983 1.8
-1.229 -0.665 2.2
-1.388 -0.7511 2.6
-1.591 -0.8608 3.0
-1.846 -0.999 3.4
-2.165 -1.172 3.8
-2.562 -1.386 4.2
-3.051 -1.651 4.6
-3.652 -1.976 5.0
-4.386 -2.373 5.4

-3.378 -3.669 -5.4
-2.812 -3.055 -5.0
-2.35 -2.552 -4.6
-1.973 -2.143 -4.2
-1.667 -1.811 -3.8
-1.422 -1.544 -3.4
-1.225 -1.331 -3.0
-1.069 -1.161 -2.6
-0.9462 -1.028 -2.2
-0.8514 -0.9249 -1.8
-0.78 -0.8473 -1.4
-0.7288 -0.7917 -1.0
-0.6956 -0.7557 -0.6
-0.6793 -0.7379 -0.2
-0.6793 -0.7379 0.2
-0.6956 -0.7557 0.6
-0.7288 -0.7917 1.0
-0.78 -0.8473 1.4
-0.8514 -0.9249 1.8
-0.9462 -1.028 2.2
-1.069 -1.161 2.6
-1.225 -1.331 3.0
-1.422 -1.544 3.4
-1.667 -1.811 3.8
-1.973 -2.143 4.2
-2.35 -2.552 4.6
-2.812 -3.055 5.0
-3.378 -3.669 5.4

-2.003 -4.567 -5.4
-1.668 -3.803 -5.0
-1.394 -3.177 -4.6
-1.17 -2.667 -4.2
-0.989 -2.255 -3.8
-0.8432 -1.922 -3.4
-0.7265 -1.656 -3.0
-0.6339 -1.445 -2.6
-0.5612 -1.279 -2.2
-0.505 -1.151 -1.8
-0.4626 -1.055 -1.4
-0.4322 -0.9854 -1.0
-0.4126 -0.9406 -0.6
-0.4029 -0.9185 -0.2
-0.4029 -0.9185 0.2
-0.4126 -0.9406 0.6
-0.4322 -0.9854 1.0
-0.4626 -1.055 1.4
-0.505 -1.151 1.8
-0.5612 -1.279 2.2
-0.6339 -1.445 2.6
-0.7265 -1.656 3.0
-0.8432 -1.922 3.4
-0.989 -2.255 3.8
-1.17 -2.667 4.2
-1.394 -3.177 4.6
-1.668 -3.803 5.0
-2.003 -4.567 5.4

-0.4118 -4.97 -5.4
-0.3429 -4.138 -5.0
-0.2865 -3.457 -4.6
-0.2405 -2.903 -4.2
-0.2033 -2.454 -3.8
-0.1733 -2.092 -3.4
-0.1494 -1.802 -3.0
-0.1303 -1.573 -2.6
-0.1154 -1.392 -2.2
-0.1038 -1.253 -1.8
-0.0951 -1.148 -1.4
-0.08886 -1.072 -1.0
-0.08482 -1.024 -0.6
-0.08283 -0.9996 -0.2
-0.08283 -0.9996 0.2
-0.08482 -1.024 0.6
-0.08886 -1.072 1.0
-0.0951 -1.148 1.4
-0.1038 -1.253 1.8
-0.1154 -1.392 2.2
-0.1303 -1.573 2.6
-0.1494 -1.802 3.0
-0.1733 -2.092 3.4
-0.2033 -2.454 3.8
-0.2405 -2.903 4.2
-0.2865 -3.457 4.6
-0.3429 -4.138 5.0
-0.4118 -4.97 5.4

1.224 -4.834 -5.4
1.019 -4.025 -5.0
0.8516 -3.363 -4.6
0.715 -2.824 -4.2
0.6044 -2.387 -3.8
0.5153 -2.035 -3.4
0.444 -1.753 -3.0
0.3874 -1.53 -2.6
0.343 -1.354 -2.2
0.3086 -1.219 -1.8
0.2827 -1.116 -1.4
0.2642 -1.043 -1.0
0.2521 -0.9957 -0.6
0.2462 -0.9723 -0.2
0.2462 -0.9723 0.2
0.2521 -0.9957 0.6
0.2642 -1.043 1.0
0.2827 -1.116 1.4
0.3086 -1.219 1.8
0.343 -1.354 2.2
0.3874 -1.53 2.6
0.444 -1.753 3.0
0.5153 -2.035 3.4
0.6044 -2.387 3.8
0.715 -2.824 4.2
0.8516 -3.363 4.6
1.019 -4.025 5.0
1.224 -4.834 5.4

2.728 -4.175 -5.4
2.271 -3.476 -5.0
1.897 -2.904 -4.6
1.593 -2.438 -4.2
1.347 -2.061 -3.8
1.148 -1.757 -3.4
0.9892 -1.514 -3.0
0.8632 -1.321 -2.6
0.7641 -1.17 -2.2
0.6876 -1.052 -1.8
0.6299 -0.9641 -1.4
0.5886 -0.9008 -1.0
0.5618 -0.8599 -0.6
0.5486 -0.8397 -0.2
0.5486 -0.8397 0.2
0.5618 -0.8599 0.6
0.5886 -0.9008 1.0
0.6299 -0.9641 1.4
0.6876 -1.052 1.8
0.7641 -1.17 2.2
0.8632 -1.321 2.6
0.9892 -1.514 3.0
1.148 -1.757 3.4
1.347 -2.061 3.8
1.593 -2.438 4.2
1.897 -2.904 4.6
2.271 -3.476 5.0
2.728 -4.175 5.4

3.935 -3.063 -5.4
3.277 -2.55 -5.0
2.738 -2.131 -4.6
2.299 -1.789 -4.2
1.943 -1.512 -3.8
1.656 -1.289 -3.4
1.427 -1.111 -3.0
1.245 -0.9693 -2.6
1.103 -0.8581 -2.2
0.992 -0.7721 -1.8
0.9088 -0.7074 -1.4
0.8492 -0.6609 -1.0
0.8105 -0.6309 -0.6
0.7915 -0.6161 -0.2
0.7915 -0.6161 0.2
0.8105 -0.6309 0.6
0.8492 -0.6609 1.0
0.9088 -0.7074 1.4
0.992 -0.7721 1.8
1.103 -0.8581 2.2
1.245 -0.9693 2.6
1.427 -1.111 3.0
1.656 -1.289 3.4
1.943 -1.512 3.8
2.299 -1.789 4.2
2.738 -2.131 4.6
3.277 -2.55 5.0
3.935 -3.063 5.4

4.717 -1.619 -5.4
3.927 -1.348 -5.0
3.281 -1.126 -4.6
2.755 -0.9457 -4.2
2.329 -0.7994 -3.8
1.985 -0.6815 -3.4
1.711 -0.5873 -3.0
1.493 -0.5124 -2.6
1.321 -0.4536 -2.2
1.189 -0.4082 -1.8
1.089 -0.3739 -1.4
1.018 -0.3494 -1.0
0.9715 -0.3335 -0.6
0.9487 -0.3257 -0.2
0.9487 -0.3257 0.2
0.9715 -0.3335 0.6
1.018 -0.3494 1.0
1.089 -0.3739 1.4
1.189 -0.4082 1.8
1.321 -0.4536 2.2
1.493 -0.5124 2.6
1.711 -0.5873 3.0
1.985 -0.6815 3.4
2.329 -0.7994 3.8
2.755 -0.9457 4.2
3.281 -1.126 4.6
3.927 -1.348 5.0
4.717 -1.619 5.4

4.987 -6.547e-21 -5.4
4.152 -5.451e-21 -5.0
3.469 -4.554e-21 -4.6
2.913 -3.824e-21 -4.2
2.462 -3.232e-21 -3.8
2.099 -2.756e-21 -3.4
1.809 -2.374e-21 -3.0
1.578 -2.072e-21 -2.6
1.397 -1.834e-21 -2.2
1.257 -1.65e-21 -1.8
1.152 -1.512e-21 -1.4
1.076 -1.413e-21 -1.0
1.027 -1.348e-21 -0.6
1.003 -1.317e-21 -0.2
1.003 -1.317e-21 0.2
1.027 -1.348e-21 0.6
1.076 -1.413e-21 1.0
1.152 -1.512e-21 1.4
1.257 -1.65e-21 1.8
1.397 -1.834e-21 2.2
1.578 -2.072e-21 2.6
1.809 -2.374e-21 3.0
2.099 -2.756e-21 3.4
2.462 -3.232e-21 3.8
2.913 -3.824e-21 4.2
3.469 -4.554e-21 4.6
4.152 -5.451e-21 5.0
4.987 -6.547e-21 5.4
};
\end{axis}
\end{tikzpicture}
\caption{$\alpha=2.2$}
\end{subfigure}
\caption{\label{fig:embedding}The embedding surface of revolution, generated by rotating the profile $\varrho(v)$ about the $z$-axis, for $\varrho_{0}=a=1$, $\kappa=0.3$, at $\alpha=1.3$ (left) and $\alpha=2.2$ (right). Both surfaces share an identical throat curvature $\varrho''(0)=2\kappa$; the visibly different flare rates reflect the deformation-onset scale of Eq.~\eqref{2.12}.}
\end{figure}
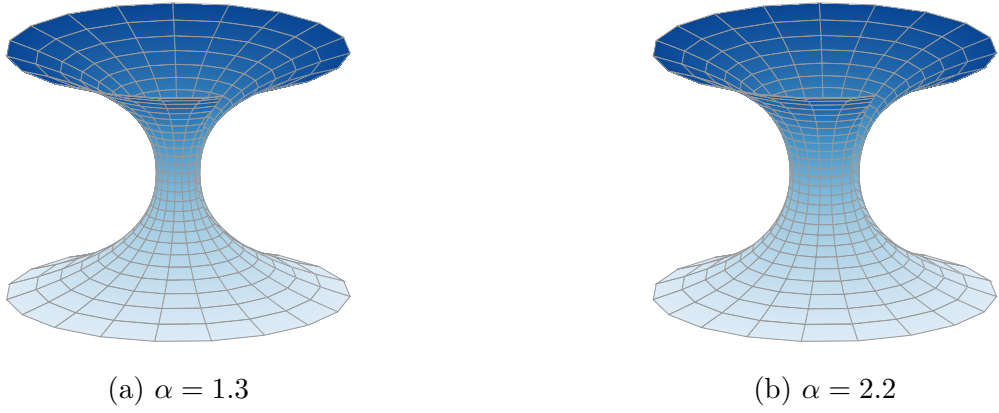

Because the argument of Eq.~\eqref{2.4} is taken as $|v|^{2n\alpha}$ rather than $v^{2n\alpha}$, the present construction is treated as a real-variable analysis throughout, without invoking analytic continuation into the complex $v$--plane; the differentiability properties established below are correspondingly formulated on the real line. In particular, Eq.~\eqref{2.6} shows that the $m$-th derivative of $\mathfrak{S}C_{\alpha}(v)$ involves the fractional power $|v|^{2\alpha-m}$ at leading order ($n=1$), which is finite at $v=0$ only when $2\alpha-m>0$, i.e., $m<2\alpha$.

Setting $m=1$ requires $\alpha>1/2$ for the first derivative to vanish at the throat, and setting $m=2$ requires $\alpha>1$ for the second derivative to vanish there as well; the admissible range $\alpha>1$ adopted throughout this work therefore secures both. This is precisely what is required by Axiom 4: since $\varrho\in C^{2}(\mathbb{R})$ demands only finiteness and continuity through the second derivative, and Eqs.~\eqref{2.7} and \eqref{2.11a} together guarantee $\varrho''(0)=2\kappa$, a finite and strictly positive quantity, the regularity requirement of Axiom 4 is satisfied throughout the admissible range $\alpha>1$, independently of $\kappa$. We note that $\alpha>1$ does not by itself guarantee finiteness of derivatives beyond second order; the third derivative, for instance, remains finite at $v=0$ only for $\alpha\geq3/2$. Since no quantity computed in this work requires more than second-order differentiability of $\varrho(v)$ at the throat -- the shape function derivative $b''(r)$ of Sec.~\ref{sec:level3} is explicitly restricted to $\varrho'(v)\neq0$, and the thin-shell formalism of Sec.~\ref{sec:level6} does not require $B_i''$ explicitly -- Axiom 4 is stated, and required, only at the order actually used in the subsequent analysis.

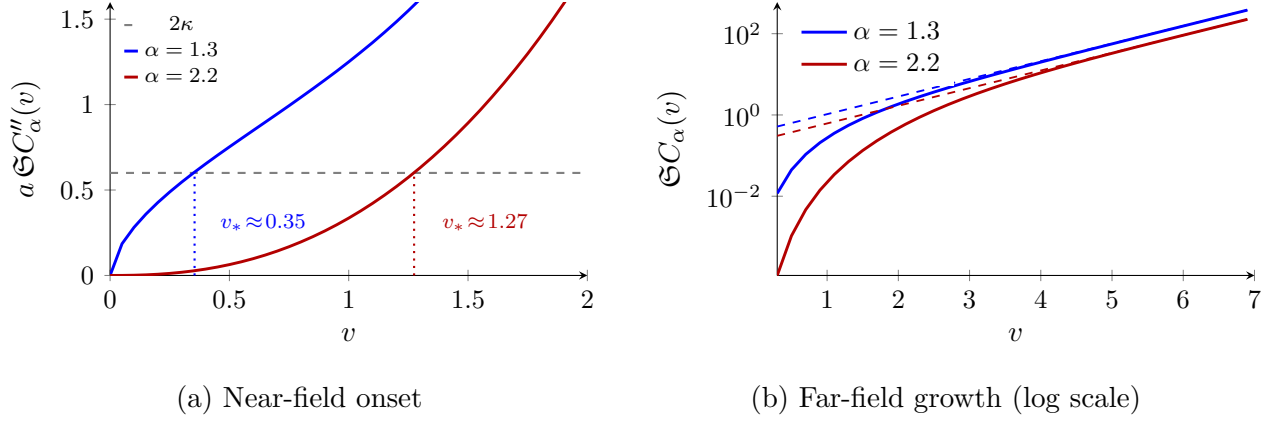
\begin{figure}
\centering
\begin{subfigure}{0.48\linewidth}
\centering
\begin{tikzpicture}
\begin{axis}[
    width=\linewidth, height=5.2cm,
    xlabel={$v$}, ylabel={$a\,\mathfrak{S}C_\alpha''(v)$},
    xmin=0, xmax=2, ymin=0, ymax=1.6,
    axis lines=left,
    legend style={font=\scriptsize, draw=none, fill=none,
                  at={(0.02,0.98)}, anchor=north west,
                  inner sep=1pt, row sep=-1pt,
                  /tikz/every even column/.append style={column sep=2pt}},
    legend image code/.code={\draw[mark repeat=2,mark phase=2,#1]
                             plot coordinates {(0cm,0cm) (0.18cm,0cm)};},
    label style={font=\small}, tick label style={font=\footnotesize},
    every axis plot/.append style={line width=1.1pt},
]
\addplot[gray, dashed, line width=0.8pt, domain=0:2] {0.6};
\addlegendentry{$2\kappa$}
\addplot[blue, solid] coordinates {(0.0,0.0) (0.05,0.1855) (0.1,0.2812) (0.15,0.3589) (0.2,0.4269) (0.25,0.4887) (0.3,0.5462) (0.35,0.6006) (0.4,0.6527) (0.45,0.7032) (0.5,0.7524) (0.55,0.8009) (0.6,0.849) (0.65,0.8969) (0.7,0.945) (0.75,0.9935) (0.8,1.043) (0.85,1.093) (0.9,1.144) (0.95,1.196) (1.0,1.25) (1.05,1.306) (1.1,1.363) (1.15,1.423) (1.2,1.485) (1.25,1.55) (1.3,1.618) (1.35,1.688) (1.4,1.762) (1.45,1.84) (1.5,1.921) (1.55,2.006) (1.6,2.095) (1.65,2.189) (1.7,2.288) (1.75,2.391) (1.8,2.5) (1.85,2.615) (1.9,2.736) (1.95,2.863) (2.0,2.997)};
\addlegendentry{$\alpha=1.3$}
\addplot[red!70!black, solid] coordinates {(0.0,0.0) (0.05,0.000253) (0.1,0.001335) (0.15,0.003534) (0.2,0.007048) (0.25,0.01204) (0.3,0.01865) (0.35,0.027) (0.4,0.0372) (0.45,0.04935) (0.5,0.06356) (0.55,0.07989) (0.6,0.09845) (0.65,0.1193) (0.7,0.1425) (0.75,0.1682) (0.8,0.1964) (0.85,0.2272) (0.9,0.2606) (0.95,0.2968) (1.0,0.3357) (1.05,0.3775) (1.1,0.4222) (1.15,0.4699) (1.2,0.5206) (1.25,0.5744) (1.3,0.6314) (1.35,0.6916) (1.4,0.7551) (1.45,0.822) (1.5,0.8923) (1.55,0.9661) (1.6,1.044) (1.65,1.125) (1.7,1.21) (1.75,1.298) (1.8,1.391) (1.85,1.488) (1.9,1.589) (1.95,1.694) (2.0,1.803)};
\addlegendentry{$\alpha=2.2$}
\draw[blue,dotted,thick] (axis cs:0.354,0) -- (axis cs:0.354,0.6);
\draw[red!70!black,dotted,thick] (axis cs:1.274,0) -- (axis cs:1.274,0.6);
\node[blue,font=\scriptsize,anchor=west] at (axis cs:0.42,0.31) {$v_*\!\approx\!0.35$};
\node[red!70!black,font=\scriptsize,anchor=west] at (axis cs:1.35,0.31) {$v_*\!\approx\!1.27$};
\end{axis}
\end{tikzpicture}
\caption{Near-field onset}
\end{subfigure}
\hfill
\begin{subfigure}{0.48\linewidth}
\centering
\begin{tikzpicture}
\begin{axis}[
    width=\linewidth, height=5.2cm,
    xlabel={$v$}, ylabel={$\mathfrak{S}C_\alpha(v)$},
    xmin=0.3, xmax=7, ymax=600,
    ymode=log,
    axis lines=left,
    legend style={font=\footnotesize, draw=none, at={(0.03,0.97)}, anchor=north west},
    label style={font=\small}, tick label style={font=\footnotesize},
    every axis plot/.append style={line width=1.1pt},
]
\addplot[blue, solid] coordinates {(0.3,0.011769) (0.5,0.044535) (0.7,0.10736) (0.9,0.208) (1.1,0.35446) (1.3,0.55557) (1.5,0.82156) (1.7,1.1646) (1.9,1.5994) (2.1,2.144) (2.3,2.8204) (2.5,3.6559) (2.7,4.6836) (2.9,5.9445) (3.1,7.4888) (3.3,9.3781) (3.5,11.688) (3.7,14.51) (3.9,17.957) (4.1,22.167) (4.3,27.309) (4.5,33.589) (4.7,41.258) (4.9,50.625) (5.1,62.064) (5.3,76.034) (5.5,93.097) (5.7,113.94) (5.9,139.39) (6.1,170.48) (6.3,208.44) (6.5,254.82) (6.7,311.46) (6.9,380.64)};
\addlegendentry{$\alpha=1.3$}
\addplot[blue, dashed, line width=0.7pt, forget plot] coordinates {(0.3,0.51918) (0.5,0.63412) (0.7,0.77452) (0.9,0.946) (1.1,1.1554) (1.3,1.4113) (1.5,1.7237) (1.7,2.1054) (1.9,2.5715) (2.1,3.1408) (2.3,3.8362) (2.5,4.6856) (2.7,5.723) (2.9,6.9901) (3.1,8.5377) (3.3,10.428) (3.5,12.737) (3.7,15.557) (3.9,19.001) (4.1,23.208) (4.3,28.346) (4.5,34.622) (4.7,42.287) (4.9,51.65) (5.1,63.085) (5.3,77.053) (5.5,94.112) (5.7,114.95) (5.9,140.4) (6.1,171.48) (6.3,209.45) (6.5,255.82) (6.7,312.46) (6.9,381.64)};
\addplot[red!70!black, solid] coordinates {(0.3,0.0001122) (0.5,0.0010621) (0.7,0.0046679) (0.9,0.014106) (1.1,0.034114) (1.3,0.071169) (1.5,0.13365) (1.7,0.23201) (1.9,0.37896) (2.1,0.58969) (2.3,0.88213) (2.5,1.2773) (2.7,1.7999) (2.9,2.4786) (3.1,3.3471) (3.3,4.4452) (3.5,5.8197) (3.7,7.5265) (3.9,9.6324) (4.1,12.218) (4.3,15.38) (4.5,19.237) (4.7,23.933) (4.9,29.644) (5.1,36.586) (5.3,45.021) (5.5,55.275) (5.7,67.743) (5.9,82.913) (6.1,101.38) (6.3,123.88) (6.5,151.3) (6.7,184.75) (6.9,225.55)};
\addlegendentry{$\alpha=2.2$}
\addplot[red!70!black, dashed, line width=0.7pt, forget plot] coordinates {(0.3,0.30679) (0.5,0.37471) (0.7,0.45767) (0.9,0.559) (1.1,0.68277) (1.3,0.83393) (1.5,1.0186) (1.7,1.2441) (1.9,1.5195) (2.1,1.8559) (2.3,2.2669) (2.5,2.7687) (2.7,3.3818) (2.9,4.1305) (3.1,5.045) (3.3,6.162) (3.5,7.5262) (3.7,9.1926) (3.9,11.228) (4.1,13.714) (4.3,16.75) (4.5,20.458) (4.7,24.988) (4.9,30.52) (5.1,37.278) (5.3,45.531) (5.5,55.612) (5.7,67.924) (5.9,82.963) (6.1,101.33) (6.3,123.77) (6.5,151.17) (6.7,184.64) (6.9,225.52)};
\end{axis}
\end{tikzpicture}
\caption{Far-field growth (log scale)}
\end{subfigure}
\caption{\label{fig:onsetfarfield}(a) The leading Mittag--Leffler curvature contribution $a\,\mathfrak{S}C_\alpha''(v)$ against $v$, crossing the base curvature $2\kappa$ (dashed) at the onset radius $v_*$ of Eq.~\eqref{2.12}, marked for each $\alpha$. (b) $\mathfrak{S}C_\alpha(v)$ (solid) on a logarithmic scale against the asymptotic form $(2\alpha)^{-1}e^{v}$ of Eq.~\eqref{2.14} (dashed): the curves become parallel at large $v$, confirming the universal exponential rate, with $\alpha$ setting only the amplitude.}
\end{figure}

Furthermore, for $v>0$, both contributions to Eq.~\eqref{2.11} are strictly positive, so that $\varrho'(v)>0$ and the embedding function $\varrho(v)$ is monotonically increasing on the positive branch (Axiom 5). Consequently, each radial coordinate
\[
r=\varrho(v)
\]
admits a locally unique correspondence with the parameter $v$, allowing the geometric quantities
\[
b(r), \qquad b^{\prime}(r), \qquad b^{\prime\prime}(r)
\]
to be expressed consistently as functions of the radial coordinate wherever the inverse mapping exists.

The radial scale at which the Mittag--Leffler deformation becomes significant relative to the local throat curvature can be quantified explicitly. Comparing the two contributions to Eq.~\eqref{2.11a} -- the constant term $2\kappa$ against the leading term of Eq.~\eqref{2.8} -- the two become comparable at
\begin{equation}\label{2.12}
v_{*}
=
\left[
\frac{2\kappa\,\Gamma(2\alpha-1)}{a}
\right]^{\frac{1}{2\alpha-2}}.
\end{equation}
Equation~\eqref{2.12} is a leading-order estimate, obtained by retaining only the $n=1$ term of the series defining $\mathfrak{S}C_{\alpha}^{\prime\prime}(v)$; it is self-consistent provided the $n=2$ term remains small in comparison at $v=v_{*}$, i.e., provided
\begin{equation}\label{2.13}
\mathcal{E}
:=
v_{*}^{2}\cdot
\frac{2\kappa\,\bigl[\Gamma(2\alpha-1)\bigr]^{2}}
{a\,\Gamma(4\alpha-1)}
\ll
1.
\end{equation}
When Eq.~\eqref{2.13} holds, Eq.~\eqref{2.12} accurately locates the radius beyond which the Mittag--Leffler deformation, rather than the local quadratic term, governs the throat curvature. For parameter choices with $\kappa/a$ of order unity or larger, Eq.~\eqref{2.13} may fail, and the true onset radius must instead be determined from the full series in Eq.~\eqref{2.6} or evaluated numerically.

\begin{figure}
\centering
\begin{subfigure}{0.48\linewidth}
\centering
\begin{tikzpicture}
\begin{axis}[
    width=\linewidth, height=6.5cm,
    view={-40}{22},
    xlabel={$\alpha$}, ylabel={$v$}, zlabel={$a\,\mathfrak{S}C_\alpha''(v)$},
    colormap/cool,
    label style={font=\scriptsize}, tick label style={font=\tiny},
    zmax=1.8,
]
\addplot3[surf, shader=interp, mesh/rows=22] table {
0.0 1.05 0.0
0.08 1.05 0.81812
0.16 1.05 0.88246
0.24 1.05 0.92926
0.32 1.05 0.97177
0.4 1.05 1.0146
0.48 1.05 1.0599
0.56 1.05 1.1093
0.64 1.05 1.1637
0.72 1.05 1.2239
0.8 1.05 1.2907
0.88 1.05 1.3646
0.96 1.05 1.4465
1.04 1.05 1.5369
1.12 1.05 1.6366
1.2 1.05 1.7462
1.28 1.05 1.8667
1.36 1.05 1.9987
1.44 1.05 2.1432
1.52 1.05 2.3012
1.6 1.05 2.4737
1.68 1.05 2.6619
1.76 1.05 2.8669
1.84 1.05 3.0902
1.92 1.05 3.3332
2.0 1.05 3.5974

0.0 1.119 0.0
0.08 1.119 0.60357
0.16 1.119 0.71447
0.24 1.119 0.79228
0.32 1.119 0.85722
0.4 1.119 0.91667
0.48 1.119 0.97438
0.56 1.119 1.0327
0.64 1.119 1.0932
0.72 1.119 1.1572
0.8 1.119 1.2258
0.88 1.119 1.2997
0.96 1.119 1.38
1.04 1.119 1.4674
1.12 1.119 1.5627
1.2 1.119 1.6667
1.28 1.119 1.7803
1.36 1.119 1.9044
1.44 1.119 2.0399
1.52 1.119 2.1877
1.6 1.119 2.3489
1.68 1.119 2.5246
1.76 1.119 2.716
1.84 1.119 2.9244
1.92 1.119 3.1512
2.0 1.119 3.3979

0.0 1.1881 0.0
0.08 1.1881 0.43526
0.16 1.1881 0.56611
0.24 1.1881 0.66214
0.32 1.1881 0.74269
0.4 1.1881 0.81521
0.48 1.1881 0.88369
0.56 1.1881 0.95065
0.64 1.1881 1.0179
0.72 1.1881 1.0868
0.8 1.1881 1.1586
0.88 1.1881 1.2343
0.96 1.1881 1.3149
1.04 1.1881 1.4012
1.12 1.1881 1.4942
1.2 1.1881 1.5947
1.28 1.1881 1.7035
1.36 1.1881 1.8217
1.44 1.1881 1.9501
1.52 1.1881 2.0897
1.6 1.1881 2.2416
1.68 1.1881 2.4068
1.76 1.1881 2.5866
1.84 1.1881 2.7822
1.92 1.1881 2.995
2.0 1.1881 3.2264

0.0 1.2571 0.0
0.08 1.2571 0.30773
0.16 1.2571 0.44004
0.24 1.2571 0.54344
0.32 1.2571 0.63272
0.4 1.2571 0.71398
0.48 1.2571 0.7906
0.56 1.2571 0.86482
0.64 1.2571 0.9383
0.72 1.2571 1.0124
0.8 1.2571 1.0882
0.88 1.2571 1.1667
0.96 1.2571 1.2489
1.04 1.2571 1.3357
1.12 1.2571 1.4281
1.2 1.2571 1.5267
1.28 1.2571 1.6327
1.36 1.2571 1.7468
1.44 1.2571 1.8701
1.52 1.2571 2.0034
1.6 1.2571 2.148
1.68 1.2571 2.3047
1.76 1.2571 2.4749
1.84 1.2571 2.6597
1.92 1.2571 2.8604
2.0 1.2571 3.0785

0.0 1.3262 0.0
0.08 1.3262 0.21378
0.16 1.3262 0.33624
0.24 1.3262 0.43872
0.32 1.3262 0.53067
0.4 1.3262 0.61626
0.48 1.3262 0.69792
0.56 1.3262 0.77732
0.64 1.3262 0.85579
0.72 1.3262 0.9344
0.8 1.3262 1.0141
0.88 1.3262 1.0959
0.96 1.3262 1.1806
1.04 1.3262 1.2689
1.12 1.3262 1.3619
1.2 1.3262 1.4603
1.28 1.3262 1.5649
1.36 1.3262 1.6767
1.44 1.3262 1.7966
1.52 1.3262 1.9255
1.6 1.3262 2.0645
1.68 1.3262 2.2147
1.76 1.3262 2.3771
1.84 1.3262 2.553
1.92 1.3262 2.7436
2.0 1.3262 2.9504

0.0 1.3952 0.0
0.08 1.3952 0.14621
0.16 1.3952 0.25299
0.24 1.3952 0.34889
0.32 1.3952 0.43869
0.4 1.3952 0.52465
0.48 1.3952 0.6082
0.56 1.3952 0.69039
0.64 1.3952 0.7721
0.72 1.3952 0.85411
0.8 1.3952 0.93716
0.88 1.3952 1.022
0.96 1.3952 1.1092
1.04 1.3952 1.1997
1.12 1.3952 1.2941
1.2 1.3952 1.3931
1.28 1.3952 1.4977
1.36 1.3952 1.6085
1.44 1.3952 1.7266
1.52 1.3952 1.8527
1.6 1.3952 1.9879
1.68 1.3952 2.1333
1.76 1.3952 2.2898
1.84 1.3952 2.4587
1.92 1.3952 2.6412
2.0 1.3952 2.8386

0.0 1.4643 0.0
0.08 1.4643 0.098592
0.16 1.4643 0.1877
0.24 1.4643 0.27366
0.32 1.4643 0.35782
0.4 1.4643 0.44092
0.48 1.4643 0.52352
0.56 1.4643 0.60608
0.64 1.4643 0.68908
0.72 1.4643 0.77298
0.8 1.4643 0.85824
0.88 1.4643 0.94539
0.96 1.4643 1.0349
1.04 1.4643 1.1275
1.12 1.4643 1.2236
1.2 1.4643 1.324
1.28 1.4643 1.4292
1.36 1.4643 1.5402
1.44 1.4643 1.6576
1.52 1.4643 1.7824
1.6 1.4643 1.9153
1.68 1.4643 2.0574
1.76 1.4643 2.2098
1.84 1.4643 2.3735
1.92 1.4643 2.5496
2.0 1.4643 2.7396

0.0 1.5333 0.0
0.08 1.5333 0.065632
0.16 1.5333 0.13749
0.24 1.5333 0.21195
0.32 1.5333 0.28825
0.4 1.5333 0.36609
0.48 1.5333 0.44539
0.56 1.5333 0.52615
0.64 1.5333 0.6085
0.72 1.5333 0.69262
0.8 1.5333 0.77875
0.88 1.5333 0.86719
0.96 1.5333 0.9583
1.04 1.5333 1.0525
1.12 1.5333 1.1502
1.2 1.5333 1.252
1.28 1.5333 1.3584
1.36 1.5333 1.4701
1.44 1.5333 1.5878
1.52 1.5333 1.7122
1.6 1.5333 1.8441
1.68 1.5333 1.9844
1.76 1.5333 2.1341
1.84 1.5333 2.2942
1.92 1.5333 2.4658
2.0 1.5333 2.6501

0.0 1.6024 0.0
0.08 1.6024 0.043177
0.16 1.6024 0.099529
0.24 1.6024 0.16225
0.32 1.6024 0.22954
0.4 1.6024 0.30054
0.48 1.6024 0.37475
0.56 1.6024 0.4519
0.64 1.6024 0.53185
0.72 1.6024 0.61457
0.8 1.6024 0.70012
0.88 1.6024 0.78863
0.96 1.6024 0.88029
1.04 1.6024 0.97538
1.12 1.6024 1.0742
1.2 1.6024 1.1772
1.28 1.6024 1.2848
1.36 1.6024 1.3974
1.44 1.6024 1.5158
1.52 1.6024 1.6405
1.6 1.6024 1.7722
1.68 1.6024 1.9118
1.76 1.6024 2.06
1.84 1.6024 2.218
1.92 1.6024 2.3866
2.0 1.6024 2.567

0.0 1.6714 0.0
0.08 1.6714 0.028097
0.16 1.6714 0.071272
0.24 1.6714 0.12286
0.32 1.6714 0.18084
0.4 1.6714 0.24413
0.48 1.6714 0.31206
0.56 1.6714 0.3842
0.64 1.6714 0.46029
0.72 1.6714 0.54015
0.8 1.6714 0.62372
0.88 1.6714 0.711
0.96 1.6714 0.80207
1.04 1.6714 0.89709
1.12 1.6714 0.99624
1.2 1.6714 1.0998
1.28 1.6714 1.2082
1.36 1.6714 1.3217
1.44 1.6714 1.4408
1.52 1.6714 1.5661
1.6 1.6714 1.6982
1.68 1.6714 1.8378
1.76 1.6714 1.9856
1.84 1.6714 2.1424
1.92 1.6714 2.3093
2.0 1.6714 2.4873

0.0 1.7405 0.0
0.08 1.7405 0.0181
0.16 1.7405 0.050523
0.24 1.7405 0.092108
0.32 1.7405 0.14105
0.4 1.7405 0.19635
0.48 1.7405 0.25732
0.56 1.7405 0.32351
0.64 1.7405 0.39462
0.72 1.7405 0.4704
0.8 1.7405 0.55074
0.88 1.7405 0.63556
0.96 1.7405 0.72487
1.04 1.7405 0.81873
1.12 1.7405 0.91726
1.2 1.7405 1.0206
1.28 1.7405 1.1291
1.36 1.7405 1.243
1.44 1.7405 1.3627
1.52 1.7405 1.4886
1.6 1.7405 1.6212
1.68 1.7405 1.7611
1.76 1.7405 1.909
1.84 1.7405 2.0656
1.92 1.7405 2.2317
2.0 1.7405 2.4083

0.0 1.8095 0.0
0.08 1.8095 0.01155
0.16 1.8095 0.035479
0.24 1.8095 0.068404
0.32 1.8095 0.10899
0.4 1.8095 0.15645
0.48 1.8095 0.21023
0.56 1.8095 0.26994
0.64 1.8095 0.33529
0.72 1.8095 0.40607
0.8 1.8095 0.48214
0.88 1.8095 0.56342
0.96 1.8095 0.64986
1.04 1.8095 0.74148
1.12 1.8095 0.83836
1.2 1.8095 0.94062
1.28 1.8095 1.0484
1.36 1.8095 1.162
1.44 1.8095 1.2817
1.52 1.8095 1.4078
1.6 1.8095 1.5408
1.68 1.8095 1.681
1.76 1.8095 1.8292
1.84 1.8095 1.986
1.92 1.8095 2.152
2.0 1.8095 2.3281

0.0 1.8786 0.0
0.08 1.8786 0.0073054
0.16 1.8786 0.024694
0.24 1.8786 0.050353
0.32 1.8786 0.08348
0.4 1.8786 0.12357
0.48 1.8786 0.17026
0.56 1.8786 0.22329
0.64 1.8786 0.28245
0.72 1.8786 0.3476
0.8 1.8786 0.41861
0.88 1.8786 0.49544
0.96 1.8786 0.57804
1.04 1.8786 0.66642
1.12 1.8786 0.76065
1.2 1.8786 0.8608
1.28 1.8786 0.96702
1.36 1.8786 1.0795
1.44 1.8786 1.1984
1.52 1.8786 1.3242
1.6 1.8786 1.457
1.68 1.8786 1.5974
1.76 1.8786 1.7458
1.84 1.8786 1.9028
1.92 1.8786 2.069
2.0 1.8786 2.2451

0.0 1.9476 0.0
0.08 1.9476 0.0045824
0.16 1.9476 0.017046
0.24 1.9476 0.036759
0.32 1.9476 0.06341
0.4 1.9476 0.096795
0.48 1.9476 0.13676
0.56 1.9476 0.1832
0.64 1.9476 0.23602
0.72 1.9476 0.29516
0.8 1.9476 0.36059
0.88 1.9476 0.43228
0.96 1.9476 0.51025
1.04 1.9476 0.59452
1.12 1.9476 0.68516
1.2 1.9476 0.78226
1.28 1.9476 0.88595
1.36 1.9476 0.99638
1.44 1.9476 1.1138
1.52 1.9476 1.2384
1.6 1.9476 1.3704
1.68 1.9476 1.5104
1.76 1.9476 1.6586
1.84 1.9476 1.8156
1.92 1.9476 1.9819
2.0 1.9476 2.1581

0.0 2.0167 0.0
0.08 2.0167 0.0028519
0.16 2.0167 0.011674
0.24 2.0167 0.026625
0.32 2.0167 0.04779
0.4 2.0167 0.075231
0.48 2.0167 0.109
0.56 2.0167 0.14914
0.64 2.0167 0.1957
0.72 2.0167 0.24872
0.8 2.0167 0.30826
0.88 2.0167 0.37436
0.96 2.0167 0.44709
1.04 2.0167 0.52654
1.12 2.0167 0.61281
1.2 2.0167 0.706
1.28 2.0167 0.80626
1.36 2.0167 0.91375
1.44 2.0167 1.0287
1.52 2.0167 1.1513
1.6 2.0167 1.2818
1.68 2.0167 1.4206
1.76 2.0167 1.568
1.84 2.0167 1.7245
1.92 2.0167 1.8905
2.0 2.0167 2.0667

0.0 2.0857 0.0
0.08 2.0857 0.0017617
0.16 2.0857 0.0079361
0.24 2.0857 0.019142
0.32 2.0857 0.03575
0.4 2.0857 0.058039
0.48 2.0857 0.086233
0.56 2.0857 0.12052
0.64 2.0857 0.16108
0.72 2.0857 0.20806
0.8 2.0857 0.26161
0.88 2.0857 0.32187
0.96 2.0857 0.38898
1.04 2.0857 0.46309
1.12 2.0857 0.54433
1.2 2.0857 0.63288
1.28 2.0857 0.72891
1.36 2.0857 0.8326
1.44 2.0857 0.94418
1.52 2.0857 1.0639
1.6 2.0857 1.192
1.68 2.0857 1.3288
1.76 2.0857 1.4747
1.84 2.0857 1.63
1.92 2.0857 1.7952
2.0 2.0857 1.9708

0.0 2.1548 0.0
0.08 2.1548 0.0010806
0.16 2.1548 0.005357
0.24 2.1548 0.013665
0.32 2.1548 0.026556
0.4 2.1548 0.044461
0.48 2.1548 0.067743
0.56 2.1548 0.096716
0.64 2.1548 0.13166
0.72 2.1548 0.17284
0.8 2.1548 0.22049
0.88 2.1548 0.27485
0.96 2.1548 0.33612
1.04 2.1548 0.40453
1.12 2.1548 0.48029
1.2 2.1548 0.56362
1.28 2.1548 0.65474
1.36 2.1548 0.75388
1.44 2.1548 0.8613
1.52 2.1548 0.97727
1.6 2.1548 1.1021
1.68 2.1548 1.236
1.76 2.1548 1.3794
1.84 2.1548 1.5328
1.92 2.1548 1.6964
2.0 2.1548 1.8709

0.0 2.2238 0.0
0.08 2.2238 0.00065842
0.16 2.2238 0.0035918
0.24 2.2238 0.0096898
0.32 2.2238 0.019594
0.4 2.2238 0.033832
0.48 2.2238 0.052861
0.56 2.2238 0.077092
0.64 2.2238 0.1069
0.72 2.2238 0.14263
0.8 2.2238 0.18461
0.88 2.2238 0.23314
0.96 2.2238 0.28854
1.04 2.2238 0.35108
1.12 2.2238 0.42106
1.2 2.2238 0.49874
1.28 2.2238 0.58443
1.36 2.2238 0.67839
1.44 2.2238 0.78094
1.52 2.2238 0.89238
1.6 2.2238 1.013
1.68 2.2238 1.1432
1.76 2.2238 1.2834
1.84 2.2238 1.4338
1.92 2.2238 1.595
2.0 2.2238 1.7675

0.0 2.2929 0.0
0.08 2.2929 0.0003986
0.16 2.2929 0.0023929
0.24 2.2929 0.0068272
0.32 2.2929 0.014365
0.4 2.2929 0.025579
0.48 2.2929 0.040985
0.56 2.2929 0.061057
0.64 2.2929 0.086239
0.72 2.2929 0.11695
0.8 2.2929 0.15358
0.88 2.2929 0.19652
0.96 2.2929 0.24614
1.04 2.2929 0.30279
1.12 2.2929 0.36684
1.2 2.2929 0.43862
1.28 2.2929 0.51849
1.36 2.2929 0.60679
1.44 2.2929 0.70388
1.52 2.2929 0.81012
1.6 2.2929 0.92587
1.68 2.2929 1.0515
1.76 2.2929 1.1875
1.84 2.2929 1.3342
1.92 2.2929 1.4921
2.0 2.2929 1.6616

0.0 2.3619 0.0
0.08 2.3619 0.00023984
0.16 2.3619 0.0015844
0.24 2.3619 0.0047808
0.32 2.3619 0.010467
0.4 2.3619 0.019221
0.48 2.3619 0.031583
0.56 2.3619 0.048063
0.64 2.3619 0.069148
0.72 2.3619 0.095306
0.8 2.3619 0.12699
0.88 2.3619 0.16465
0.96 2.3619 0.2087
1.04 2.3619 0.25957
1.12 2.3619 0.31769
1.2 2.3619 0.38345
1.28 2.3619 0.45728
1.36 2.3619 0.53958
1.44 2.3619 0.63076
1.52 2.3619 0.73125
1.6 2.3619 0.84146
1.68 2.3619 0.96184
1.76 2.3619 1.0928
1.84 2.3619 1.2349
1.92 2.3619 1.3885
2.0 2.3619 1.5543

0.0 2.431 0.0
0.08 2.431 0.00014346
0.16 2.431 0.0010429
0.24 2.431 0.0033283
0.32 2.431 0.0075819
0.4 2.431 0.014359
0.48 2.431 0.024196
0.56 2.431 0.037613
0.64 2.431 0.05512
0.72 2.431 0.077217
0.8 2.431 0.10439
0.88 2.431 0.13714
0.96 2.431 0.17593
1.04 2.431 0.22124
1.12 2.431 0.27354
1.2 2.431 0.3333
1.28 2.431 0.401
1.36 2.431 0.47709
1.44 2.431 0.56207
1.52 2.431 0.65639
1.6 2.431 0.76055
1.68 2.431 0.87503
1.76 2.431 1.0004
1.84 2.431 1.137
1.92 2.431 1.2856
2.0 2.431 1.4466

0.0 2.5 0.0
0.08 2.5 8.5333e-5
0.16 2.5 0.00068267
0.24 2.5 0.002304
0.32 2.5 0.0054613
0.4 2.5 0.010667
0.48 2.5 0.018432
0.56 2.5 0.02927
0.64 2.5 0.043691
0.72 2.5 0.06221
0.8 2.5 0.085337
0.88 2.5 0.11359
0.96 2.5 0.14747
1.04 2.5 0.18751
1.12 2.5 0.23422
1.2 2.5 0.28811
1.28 2.5 0.3497
1.36 2.5 0.41953
1.44 2.5 0.49812
1.52 2.5 0.58601
1.6 2.5 0.68373
1.68 2.5 0.79185
1.76 2.5 0.91091
1.84 2.5 1.0415
1.92 2.5 1.1842
2.0 2.5 1.3397

};
\addplot3[surf, opacity=0.35, fill=black, mesh/rows=2, draw=none] table {
1.05 0 0.6
1.05 2 0.6

2.5 0 0.6
2.5 2 0.6
};
\node[font=\tiny] at (axis cs:1.3,2.05,0.62) {$z=2\kappa$};
\end{axis}
\end{tikzpicture}
\caption{Near field}
\end{subfigure}
\hfill
\begin{subfigure}{0.48\linewidth}
\centering
\begin{tikzpicture}
\begin{axis}[
    width=\linewidth, height=6.5cm,
    view={-40}{22},
    xlabel={$\alpha$}, ylabel={$v$}, zlabel={$\mathfrak{S}C_\alpha(v)\big/\tfrac{1}{2\alpha}e^{v}$},
    colormap/cool,
    label style={font=\scriptsize}, tick label style={font=\tiny},
]
\addplot3[surf, shader=interp, mesh/rows=20] table {
0.4 1.05 0.094438
0.66522 1.05 0.21478
0.93043 1.05 0.34311
1.1957 1.05 0.46357
1.4609 1.05 0.56926
1.7261 1.05 0.65828
1.9913 1.05 0.73127
2.2565 1.05 0.79003
2.5217 1.05 0.83673
2.787 1.05 0.8735
3.0522 1.05 0.90225
3.3174 1.05 0.92461
3.5826 1.05 0.94195
3.8478 1.05 0.95535
4.113 1.05 0.96568
4.3783 1.05 0.97364
4.6435 1.05 0.97976
4.9087 1.05 0.98447
5.1739 1.05 0.98808
5.4391 1.05 0.99086
5.7043 1.05 0.99299
5.9696 1.05 0.99462
6.2348 1.05 0.99587
6.5 1.05 0.99684

0.4 1.1263 0.075449
0.66522 1.1263 0.18441
0.93043 1.1263 0.30779
1.1957 1.1263 0.42833
1.4609 1.1263 0.5371
1.7261 1.1263 0.63055
1.9913 1.1263 0.70829
2.2565 1.1263 0.77151
2.5217 1.1263 0.8221
2.787 1.1263 0.8621
3.0522 1.1263 0.89347
3.3174 1.1263 0.9179
3.5826 1.1263 0.93684
3.8478 1.1263 0.95148
4.113 1.1263 0.96275
4.3783 1.1263 0.97143
4.6435 1.1263 0.97809
4.9087 1.1263 0.9832
5.1739 1.1263 0.98712
5.4391 1.1263 0.99013
5.7043 1.1263 0.99244
5.9696 1.1263 0.9942
6.2348 1.1263 0.99556
6.5 1.1263 0.99659

0.4 1.2026 0.05958
0.66522 1.2026 0.15675
0.93043 1.2026 0.27379
1.1957 1.2026 0.39307
1.4609 1.2026 0.50402
1.7261 1.2026 0.6015
1.9913 1.2026 0.68391
2.2565 1.2026 0.75172
2.5217 1.2026 0.80642
2.787 1.2026 0.84991
3.0522 1.2026 0.88411
3.3174 1.2026 0.91079
3.5826 1.2026 0.93148
3.8478 1.2026 0.94744
4.113 1.2026 0.95973
4.3783 1.2026 0.96916
4.6435 1.2026 0.97639
4.9087 1.2026 0.98193
5.1739 1.2026 0.98617
5.4391 1.2026 0.98941
5.7043 1.2026 0.99189
5.9696 1.2026 0.99379
6.2348 1.2026 0.99525
6.5 1.2026 0.99636

0.4 1.2789 0.04653
0.66522 1.2789 0.13191
0.93043 1.2789 0.24147
1.1957 1.2789 0.35814
1.4609 1.2789 0.47024
1.7261 1.2789 0.57116
1.9913 1.2789 0.65807
2.2565 1.2789 0.73054
2.5217 1.2789 0.78957
2.787 1.2789 0.83681
3.0522 1.2789 0.8741
3.3174 1.2789 0.90323
3.5826 1.2789 0.92583
3.8478 1.2789 0.94325
4.113 1.2789 0.95662
4.3783 1.2789 0.96687
4.6435 1.2789 0.97469
4.9087 1.2789 0.98067
5.1739 1.2789 0.98523
5.4391 1.2789 0.98871
5.7043 1.2789 0.99137
5.9696 1.2789 0.9934
6.2348 1.2789 0.99495
6.5 1.2789 0.99613

0.4 1.3553 0.035956
0.66522 1.3553 0.10994
0.93043 1.3553 0.21113
1.1957 1.3553 0.32391
1.4609 1.3553 0.43603
1.7261 1.3553 0.53967
1.9913 1.3553 0.63074
2.2565 1.3553 0.70787
2.5217 1.3553 0.77141
2.787 1.3553 0.82265
3.0522 1.3553 0.8633
3.3174 1.3553 0.89515
3.5826 1.3553 0.91985
3.8478 1.3553 0.93886
4.113 1.3553 0.95343
4.3783 1.3553 0.96455
4.6435 1.3553 0.97301
4.9087 1.3553 0.97944
5.1739 1.3553 0.98433
5.4391 1.3553 0.98805
5.7043 1.3553 0.99088
5.9696 1.3553 0.99303
6.2348 1.3553 0.99467
6.5 1.3553 0.99592

0.4 1.4316 0.027508
0.66522 1.4316 0.090776
0.93043 1.4316 0.18304
1.1957 1.4316 0.29075
1.4609 1.4316 0.40172
1.7261 1.4316 0.5072
1.9913 1.4316 0.60196
2.2565 1.4316 0.68362
2.5217 1.4316 0.75178
2.787 1.4316 0.80729
3.0522 1.4316 0.8516
3.3174 1.4316 0.88642
3.5826 1.4316 0.91346
3.8478 1.4316 0.93426
4.113 1.4316 0.95015
4.3783 1.4316 0.96222
4.6435 1.4316 0.97136
4.9087 1.4316 0.97827
5.1739 1.4316 0.98349
5.4391 1.4316 0.98744
5.7043 1.4316 0.99043
5.9696 1.4316 0.9927
6.2348 1.4316 0.99442
6.5 1.4316 0.99573

0.4 1.5079 0.020846
0.66522 1.5079 0.074273
0.93043 1.5079 0.15736
1.1957 1.5079 0.25902
1.4609 1.5079 0.36766
1.7261 1.5079 0.474
1.9913 1.5079 0.57183
2.2565 1.5079 0.65776
2.5217 1.5079 0.73057
2.787 1.5079 0.79054
3.0522 1.5079 0.83881
3.3174 1.5079 0.87693
3.5826 1.5079 0.90659
3.8478 1.5079 0.92939
4.113 1.5079 0.94675
4.3783 1.5079 0.95988
4.6435 1.5079 0.96976
4.9087 1.5079 0.97718
5.1739 1.5079 0.98275
5.4391 1.5079 0.98692
5.7043 1.5079 0.99006
5.9696 1.5079 0.99242
6.2348 1.5079 0.99421
6.5 1.5079 0.99557

0.4 1.5842 0.015655
0.66522 1.5842 0.060244
0.93043 1.5842 0.13418
1.1957 1.5842 0.22902
1.4609 1.5842 0.33422
1.7261 1.5842 0.44037
1.9913 1.5842 0.5405
2.2565 1.5842 0.6303
2.5217 1.5842 0.70768
2.787 1.5842 0.77226
3.0522 1.5842 0.82476
3.3174 1.5842 0.8665
3.5826 1.5842 0.89909
3.8478 1.5842 0.92416
4.113 1.5842 0.9432
4.3783 1.5842 0.95753
4.6435 1.5842 0.96824
4.9087 1.5842 0.97621
5.1739 1.5842 0.98212
5.4391 1.5842 0.98652
5.7043 1.5842 0.98979
5.9696 1.5842 0.99223
6.2348 1.5842 0.99407
6.5 1.5842 0.99545

0.4 1.6605 0.011657
0.66522 1.6605 0.04846
0.93043 1.6605 0.1135
1.1957 1.6605 0.20099
1.4609 1.6605 0.30176
1.7261 1.6605 0.40665
1.9913 1.6605 0.50822
2.2565 1.6605 0.60133
2.5217 1.6605 0.68306
2.787 1.6605 0.75231
3.0522 1.6605 0.80927
3.3174 1.6605 0.85495
3.5826 1.6605 0.89081
3.8478 1.6605 0.91845
4.113 1.6605 0.93942
4.3783 1.6605 0.95514
4.6435 1.6605 0.96679
4.9087 1.6605 0.97538
5.1739 1.6605 0.98167
5.4391 1.6605 0.98628
5.7043 1.6605 0.98966
5.9696 1.6605 0.99216
6.2348 1.6605 0.99401
6.5 1.6605 0.9954

0.4 1.7368 0.0086098
0.66522 1.7368 0.03867
0.93043 1.7368 0.095271
1.1957 1.7368 0.1751
1.4609 1.7368 0.27061
1.7261 1.7368 0.37321
1.9913 1.7368 0.47527
2.2565 1.7368 0.57101
2.5217 1.7368 0.65673
2.787 1.7368 0.73058
3.0522 1.7368 0.79216
3.3174 1.7368 0.84208
3.5826 1.7368 0.88156
3.8478 1.7368 0.91212
4.113 1.7368 0.93532
4.3783 1.7368 0.95265
4.6435 1.7368 0.96541
4.9087 1.7368 0.97471
5.1739 1.7368 0.98142
5.4391 1.7368 0.98626
5.7043 1.7368 0.98973
5.9696 1.7368 0.99225
6.2348 1.7368 0.99408
6.5 1.7368 0.99543

0.4 1.8132 0.00631
0.66522 1.8132 0.030623
0.93043 1.8132 0.079376
1.1957 1.8132 0.15146
1.4609 1.8132 0.24104
1.7261 1.8132 0.34039
1.9913 1.8132 0.44196
2.2565 1.8132 0.53954
2.5217 1.8132 0.62876
2.787 1.8132 0.70701
3.0522 1.8132 0.77328
3.3174 1.8132 0.82769
3.5826 1.8132 0.87113
3.8478 1.8132 0.90498
4.113 1.8132 0.93075
4.3783 1.8132 0.94998
4.6435 1.8132 0.96405
4.9087 1.8132 0.9742
5.1739 1.8132 0.98141
5.4391 1.8132 0.98649
5.7043 1.8132 0.99005
5.9696 1.8132 0.99255
6.2348 1.8132 0.99433
6.5 1.8132 0.9956

0.4 1.8895 0.0045906
0.66522 1.8895 0.024074
0.93043 1.8895 0.065661
1.1957 1.8895 0.1301
1.4609 1.8895 0.21328
1.7261 1.8895 0.30853
1.9913 1.8895 0.40865
2.2565 1.8895 0.50721
2.5217 1.8895 0.59929
2.787 1.8895 0.68163
3.0522 1.8895 0.75253
3.3174 1.8895 0.81159
3.5826 1.8895 0.85931
3.8478 1.8895 0.89681
4.113 1.8895 0.92553
4.3783 1.8895 0.94699
4.6435 1.8895 0.96265
4.9087 1.8895 0.97383
5.1739 1.8895 0.98165
5.4391 1.8895 0.98703
5.7043 1.8895 0.99068
5.9696 1.8895 0.99314
6.2348 1.8895 0.99481
6.5 1.8895 0.99595

0.4 1.9658 0.0033162
0.66522 1.9658 0.018794
0.93043 1.9658 0.053941
1.1957 1.9658 0.11101
1.4609 1.9658 0.1875
1.7261 1.9658 0.27794
1.9913 1.9658 0.37568
2.2565 1.9658 0.47433
2.5217 1.9658 0.56855
2.787 1.9658 0.6545
3.0522 1.9658 0.72986
3.3174 1.9658 0.79363
3.5826 1.9658 0.84587
3.8478 1.9658 0.88739
4.113 1.9658 0.91945
4.3783 1.9658 0.94351
4.6435 1.9658 0.96107
4.9087 1.9658 0.97354
5.1739 1.9658 0.98214
5.4391 1.9658 0.98791
5.7043 1.9658 0.99168
5.9696 1.9658 0.99409
6.2348 1.9658 0.9956
6.5 1.9658 0.99655

0.4 2.0421 0.0023795
0.66522 2.0421 0.014573
0.93043 2.0421 0.04402
1.1957 2.0421 0.0941
1.4609 2.0421 0.16379
1.7261 2.0421 0.24886
1.9913 2.0421 0.3434
2.2565 2.0421 0.44122
2.5217 2.0421 0.53678
2.787 2.0421 0.62578
3.0522 2.0421 0.70528
3.3174 2.0421 0.77371
3.5826 2.0421 0.83064
3.8478 2.0421 0.87649
4.113 2.0421 0.91227
4.3783 2.0421 0.93934
4.6435 2.0421 0.95917
4.9087 2.0421 0.97322
5.1739 2.0421 0.98283
5.4391 2.0421 0.98913
5.7043 2.0421 0.9931
5.9696 2.0421 0.99546
6.2348 2.0421 0.99679
6.5 2.0421 0.99749

0.4 2.1184 0.0016964
0.66522 2.1184 0.011228
0.93043 2.1184 0.035695
1.1957 2.1184 0.079265
1.4609 2.1184 0.1422
1.7261 2.1184 0.2215
1.9913 2.1184 0.31211
2.2565 2.1184 0.40823
2.5217 2.1184 0.5043
2.787 2.1184 0.59565
3.0522 2.1184 0.67886
3.3174 2.1184 0.75179
3.5826 2.1184 0.81346
3.8478 2.1184 0.86387
4.113 2.1184 0.90374
4.3783 2.1184 0.93423
4.6435 2.1184 0.95674
4.9087 2.1184 0.97273
5.1739 2.1184 0.98363
5.4391 2.1184 0.99067
5.7043 2.1184 0.99494
5.9696 2.1184 0.99731
6.2348 2.1184 0.99845
6.5 2.1184 0.99885

0.4 2.1947 0.001202
0.66522 2.1947 0.0085975
0.93043 2.1947 0.028766
1.1957 2.1947 0.066363
1.4609 2.1947 0.12272
1.7261 2.1947 0.196
1.9913 2.1947 0.28208
2.2565 2.1947 0.37569
2.5217 2.1947 0.47141
2.787 2.1947 0.56439
3.0522 2.1947 0.65076
3.3174 2.1947 0.72787
3.5826 2.1947 0.79423
3.8478 2.1947 0.84937
4.113 2.1947 0.89363
4.3783 2.1947 0.92793
4.6435 2.1947 0.95354
4.9087 2.1947 0.97188
5.1739 2.1947 0.9844
5.4391 2.1947 0.99245
5.7043 2.1947 0.99721
5.9696 2.1947 0.99968
6.2348 2.1947 1.0007
6.5 2.1947 1.0007

0.4 2.2711 0.00084657
0.66522 2.2711 0.0065442
0.93043 2.2711 0.023046
1.1957 2.2711 0.055235
1.4609 2.2711 0.1053
1.7261 2.2711 0.17245
1.9913 2.2711 0.25353
2.2565 2.2711 0.34392
2.5217 2.2711 0.43847
2.787 2.2711 0.53227
3.0522 2.2711 0.62117
3.3174 2.2711 0.70205
3.5826 2.2711 0.77293
3.8478 2.2711 0.83284
4.113 2.2711 0.88172
4.3783 2.2711 0.92019
4.6435 2.2711 0.94932
4.9087 2.2711 0.97044
5.1739 2.2711 0.98498
5.4391 2.2711 0.99436
5.7043 2.2711 0.99985
5.9696 2.2711 1.0026
6.2348 2.2711 1.0035
6.5 2.2711 1.0032

0.4 2.3474 0.00059284
0.66522 2.3474 0.0049527
0.93043 2.3474 0.018357
1.1957 2.3474 0.045712
1.4609 2.3474 0.089836
1.7261 2.3474 0.15089
1.9913 2.3474 0.22664
2.2565 2.3474 0.31319
2.5217 2.3474 0.40579
2.787 2.3474 0.49962
3.0522 2.3474 0.59034
3.3174 2.3474 0.67447
3.5826 2.3474 0.74956
3.8478 2.3474 0.81418
4.113 2.3474 0.86783
4.3783 2.3474 0.91077
4.6435 2.3474 0.9438
4.9087 2.3474 0.96813
5.1739 2.3474 0.98513
5.4391 2.3474 0.99622
5.7043 2.3474 1.0028
5.9696 2.3474 1.006
6.2348 2.3474 1.0069
6.5 2.3474 1.0064

0.4 2.4237 0.00041287
0.66522 2.4237 0.0037276
0.93043 2.4237 0.014542
1.1957 2.4237 0.037624
1.4609 2.4237 0.076229
1.7261 2.4237 0.13131
1.9913 2.4237 0.20153
2.2565 2.4237 0.28374
2.5217 2.4237 0.37368
2.787 2.4237 0.46675
3.0522 2.4237 0.55855
3.3174 2.4237 0.64533
3.5826 2.4237 0.72423
3.8478 2.4237 0.79338
4.113 2.4237 0.85182
4.3783 2.4237 0.89944
4.6435 2.4237 0.93674
4.9087 2.4237 0.9647
5.1739 2.4237 0.9846
5.4391 2.4237 0.99782
5.7043 2.4237 1.0058
5.9696 2.4237 1.0097
6.2348 2.4237 1.0109
6.5 2.4237 1.0102

0.4 2.5 0.000286
0.66522 2.5 0.0027907
0.93043 2.5 0.011459
1.1957 2.5 0.030802
1.4609 2.5 0.064342
1.7261 2.5 0.11368
1.9913 2.5 0.17828
2.2565 2.5 0.25576
2.5217 2.5 0.34243
2.787 2.5 0.434
3.0522 2.5 0.52612
3.3174 2.5 0.61487
3.5826 2.5 0.69709
3.8478 2.5 0.77046
4.113 2.5 0.83363
4.3783 2.5 0.88604
4.6435 2.5 0.92788
4.9087 2.5 0.95986
5.1739 2.5 0.9831
5.4391 2.5 0.99891
5.7043 2.5 1.0087
5.9696 2.5 1.0138
6.2348 2.5 1.0154
6.5 2.5 1.0148

};
\end{axis}
\end{tikzpicture}
\caption{Far field}
\end{subfigure}
\caption{\label{fig:onsetfarfields}(a) The leading Mittag--Leffler curvature contribution $a\,\mathfrak{S}C_\alpha''(v)$ as a surface over the full admissible $(\alpha,v)$ domain; the semi-transparent plane marks the base curvature $2\kappa$, and its intersection with the surface traces the onset radius $v_*(\alpha)$ of Eq.~\eqref{2.12} continuously across $\alpha$. (b) The ratio $\mathfrak{S}C_\alpha(v)/[(2\alpha)^{-1}e^{v}]$ as a surface over the same domain: it flattens to the plane $z=1$ as $v$ grows, for every $\alpha$ simultaneously, confirming the universal far-field rate of Eq.~\eqref{2.14}.}
\end{figure}
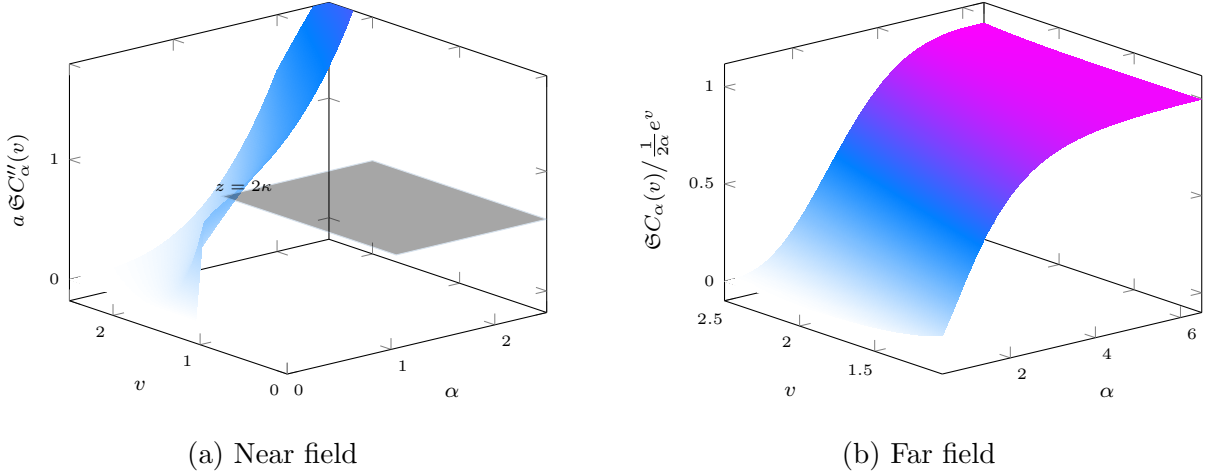

The behavior of the embedding at large $|v|$ follows from the closed-form expression in Eq.~\eqref{2.4}. Using the standard large-argument asymptotic of the Mittag--Leffler function, $E_{\rho}(z)\sim\rho^{-1}\exp(z^{1/\rho})$ as $z\rightarrow+\infty$ along the positive real axis, with $\rho=2\alpha$ and $z=|v|^{2\alpha}$, one obtains
\begin{equation}\label{2.14}
\mathfrak{S}C_{\alpha}(v)
\sim
\frac{1}{2\alpha}\,e^{|v|},
\qquad
|v|\rightarrow\infty,
\end{equation}
so that $\varrho(v)\sim a(2\alpha)^{-1}e^{|v|}$ at large $|v|$. The asymptotic growth rate is exponential with rate $1$, independently of $\alpha$; the deformation parameter $\alpha$ instead sets only the amplitude of this growth, through the prefactor $a/(2\alpha)$, delaying or advancing the onset of the exponential regime without altering its ultimate rate. This far-field behavior characterizes the embedding in isolation; the matching to the exterior vacuum solution, and the corresponding restriction of the interior geometry to a finite shell radius, is carried out in Sec.~\ref{sec:level6}.

Geometrically, the resulting embedding surface may therefore be interpreted as a fractional deformation of the classical catenoid, with the throat's local curvature fixed independently by $\kappa$ (Axiom 3), and the parameter $\alpha$ governing the onset radius, Eq.~\eqref{2.12}, and far-field amplitude, Eq.~\eqref{2.14}, of the departure from this local geometry, relative to the classical catenoidal configuration recovered as the boundary limit $\alpha\rightarrow1^{+}$, $\kappa/a\rightarrow0^{+}$.

\section{\label{sec:level3}The Metric Ansatz}

In order to transform the purely geometric framework discussed in Section \ref{sec:level2} into a comprehensive spacetime that aligns with Einstein's theory of gravitation, a general static and spherically (pseudo-spherically) symmetric line element is employed in curvature coordinates, expressed as
\begin{equation}\label{3.1}
    {ds}^{2} = -e^{2\Phi(r)}\ {dt}^{2} + \frac{{dr}^{2}}{1 - \frac{b(r)}{r}} + r^{2}\ {d\Omega_{k}}^{2}.
\end{equation}
The variable $\Phi(r)$ represents the redshift function, which is now variable rather than constant, and it governs both the gravitational potential and tidal influences \cite{morris1988wormholes, hawking2023large, thorne2000gravitation}. The function $b(r)$ represents the shape function determining the spatial structure of the wormhole geometry. Furthermore, ${d\Omega_{k}}^{2}$ denotes the metric on a two-dimensional surface of constant curvature $k = \{+1, -1\}$, where 
\begin{equation*}
    {d\Omega_{+1}}^{2} = du^{2} + \sin^{2} u \  {d\phi}^{2}, \ \ \qquad \qquad  {d\Omega_{-1}}^{2} = du^{2} + \sinh^{2} u \ {d\phi}^{2},
\end{equation*}
corresponds to spherical and pseudo-spherical topologies, respectively. In the present work, the spherical case, $k=+1$, is adopted throughout; the pseudo-spherical case is retained above only to display the general form of the ansatz.

The coordinate $r$ denotes the circumferential radius, defined such that the proper circumference around the symmetry axis is given by $2\pi r$. The metric \eqref{3.1} reduces to the standard Morris--Thorne form for suitable choices of $\Phi(r)$ and $b(r)$ \cite{morris1988wormholes}. Notably, this metric is fully consistent with the EFEs \cite{einstein1915feldgleichungen}. Such a framework facilitates a deeper understanding of the dynamic interplay between geometry and gravitation as prescribed by general relativity. In the present work, however, the shape function is not introduced phenomenologically but is derived directly from the embedding geometry developed in Section \ref{sec:level2}.

The generalized super--hyperbolic embedding, recalling Eq.~\eqref{2.4.1} of Sec.~\ref{subsec:lavel2.2}, is specified through
\begin{equation}\label{gps1}
r
\equiv
\varrho(v)
=
\varrho_{0}+\kappa v^{2}+a\,\mathfrak{S}C_{\alpha}(v),
\qquad
z=2av,
\end{equation}
where
\begin{equation}\label{gps1a}
\mathfrak{S}C_{\alpha}(v)
=
\sum_{n=1}^{\infty}
\frac{|v|^{2n\alpha}}{\Gamma(2n\alpha+1)},
\qquad
\alpha>1.
\end{equation}
Here, $\alpha > 1$ governing the embedding’s behavior away from the throat and $a>0$ denotes the geometric scaling parameter controlling the axial stretching of the embedding, $\varrho_{0}>0$ is the throat radius, and $\kappa>0$ is the local throat-curvature parameter of Sec.~\ref{subsec:lavel2.2}. The parameter $\alpha$ characterizes the higher-order deformation of the embedding away from the throat, relative to the classical catenoidal geometry recovered in the boundary limit $\alpha\rightarrow1^{+}$, $\kappa/a\rightarrow0^{+}$ (Secs.~\ref{subsec:level2.1}, \ref{subsec:lavel2.3}).

Differentiating Eq.~\eqref{gps1} with respect to the embedding coordinate $v$ yields
\begin{equation}\label{gps2}
\varrho^{\prime}(v)
=
2\kappa v+a\,\mathfrak{S}C_{\alpha}^{\prime}(v)
=
2\kappa v+a\,\mathrm{sgn}(v)\sum_{n=1}^{\infty}
\frac{|v|^{2n\alpha-1}}{\Gamma(2n\alpha)},
\end{equation}
and
\begin{equation}\label{gps3}
\varrho^{\prime\prime}(v)
=
2\kappa+a\,\mathfrak{S}C_{\alpha}^{\prime\prime}(v)
=
2\kappa+a
\sum_{n=1}^{\infty}
\frac{|v|^{2n\alpha-2}}{\Gamma(2n\alpha-1)}.
\end{equation}

For convenience, we introduce the auxiliary quantity
\begin{equation}\label{3.3}
Q(v)
:=
\varrho^{\prime}(v)^{2}
+
4a^{2}.
\end{equation}

We now consider the isometric embedding of an equatorial spatial slice into Euclidean $\mathbb{R}^{3}$ with cylindrical coordinates $(r,\theta,z)$. Using the embedding relation \eqref{gps1}, one obtains
\begin{equation}\label{3.4}
\frac{dz}{dr}
=
\frac{dz/dv}{dr/dv}
=
\frac{2a}{\varrho^{\prime}(v)},
\qquad
r=\varrho(v).
\end{equation}
The induced spatial line element on the embedded surface therefore becomes
\begin{equation}\label{3.5}
dl^{2}
=
\left[
1+
\left(\frac{dz}{dr}\right)^{2}
\right]dr^{2}
+
r^{2}d\theta^{2},
\end{equation}
which, upon substitution of Eq.~\eqref{3.4}, yields
\begin{equation}\label{3.5a}
dl^{2}
=
\frac{Q(v)}{\varrho^{\prime}(v)^{2}}
\,dr^{2}
+
r^{2}d\theta^{2}.
\end{equation}

Comparing the radial sector of Eq.~\eqref{3.5a} with the spatial part of the wormhole metric \eqref{3.1}, one obtains the embedding identity
\begin{equation}\label{3.6}
\frac{1}{1-\frac{b(r)}{r}}
=
\frac{Q(v)}{\varrho^{\prime}(v)^{2}}.
\end{equation}
Solving for the shape function gives
\begin{equation}\label{3.7}
b(r)
=
\frac{4a^{2}r}{Q(v)},
\end{equation}
where the dependence on the deformation parameter $\alpha$ enters implicitly through $\varrho(v)$ and its derivatives. 

Differentiating Eq.~\eqref{3.7} with respect to $r$ through the parameterization $r=\varrho(v)$ yields
\begin{equation}\label{3.8}
b^{\backprime}(r)
=
\frac{4a^{2}}{Q(v)}
\left[
1-
\frac{2r\,\varrho^{\prime\prime}(v)}{Q(v)}
\right].
\end{equation}
Similarly, a second differentiation gives
\begin{equation}\label{3.9}
b^{\backprime\backprime}(r)
=
-
\frac{8a^{2}r\,\varrho^{\prime\prime\prime}(v)}
{\varrho^{\prime}(v)\,Q(v)^{2}}
-
\frac{16a^{2}\varrho^{\prime\prime}(v)}
{Q(v)^{2}}
+
\frac{32a^{2}r\,[\varrho^{\prime\prime}(v)]^{2}}
{Q(v)^{3}},
\end{equation}
valid in regions where $\varrho^{\prime}(v)\neq0$.

At the throat,
\[
v=0,
\qquad
r=\varrho(0)=\varrho_{0}\equiv r_{\mathrm{th}},
\]
and
\[
\varrho^{\prime}(0)=0.
\]
Consequently,
\begin{equation}\label{3.10}
b(r_{\mathrm{th}})
=
r_{\mathrm{th}},
\end{equation}
which is the standard throat condition for traversable wormholes. Furthermore, using $\varrho''(0)=2\kappa$ from Eq.~\eqref{gps3},
\begin{equation}\label{3.10a}
b^{\backprime}(r_{\mathrm{th}})
=
1
-
\frac{r_{\mathrm{th}}\,
\varrho^{\prime\prime}(0)}
{2a^{2}}
=
1
-
\frac{\varrho_{0}\,\kappa}{a^{2}}.
\end{equation}
The flare--out condition
\[
b^{\backprime}(r_{\mathrm{th}})<1
\]
is therefore satisfied automatically, since $\varrho_{0},\kappa,a>0$ are required by Axioms 1 and 3 of Sec.~\ref{subsec:lavel2.2} -- flare-out is a direct consequence of the local curvature term $\kappa$, rather than a condition on the deformation parameter $\alpha$. Moreover, since Axiom 4 establishes $\varrho\in C^{2}(\mathbb{R})$ throughout the admissible range $\alpha>1$ (Sec.~\ref{subsec:lavel2.3}), the throat is both geometrically open and $C^{2}$-regular, without further qualification on $\alpha$.

Substituting the shape function \eqref{3.7} into the metric \eqref{3.1} yields the generalized super--hyperbolic wormhole spacetime
\begin{equation}\label{3.11}
ds^{2}
=
-e^{2\Phi(r)}dt^{2}
+
\frac{dr^{2}}
{1-\frac{4a^{2}}{Q(v)}}
+
r^{2}d\Omega_{k=+1}^{2}.
\end{equation}
Expressed entirely in terms of the embedding coordinate $v$, the metric becomes
\begin{equation}\label{3.12}
ds^{2}
=
-e^{2\Phi(v)}dt^{2}
+
Q(v)\,dv^{2}
+
\varrho(v)^{2}d\Omega_{+1}^{2},
\end{equation}
where
\[
Q(v)
=
\varrho^{\prime}(v)^{2}
+
4a^{2},
\]
and
\[
d\Omega_{+1}^{2}
=
du^{2}
+
\sin^{2}u\,d\phi^{2}.
\]

The coordinate $v$ therefore parametrizes the embedding direction of the throat geometry, while the physical circumferential radius is determined by $\varrho(v)$. The resulting spacetime is static and spherically symmetric, with the local throat curvature fixed by $\kappa$ and the parameter $\alpha$ governing the onset radius and far-field amplitude of the departure from this local geometry, relative to the classical catenoidal limit approached as $\alpha\rightarrow1^{+}$, $\kappa/a\rightarrow0^{+}$ (Sec.~\ref{subsec:lavel2.3}). It is significant to note that the constant $2a$ corresponds precisely to the axial slope $dz/dv$ defined by the embedding relation $z = 2av$.

\section{\label{sec:level4}Tidal Effects and Redshift Function}

To incorporate tidal gravitational effects in a controlled and mathematically smooth manner, we choose a one-parameter family of redshift functions that is even about the throat and regular there:
\begin{equation}\label{5.1}
\Phi(r)
=
\Phi_{0}\,
\ln\!\left[
\cosh\!\left(\frac{r-r_{\mathrm{th}}}{\Phi_{1}}\right)
\right],
\end{equation}
where $\Phi_{0}$ sets the redshift amplitude and $\Phi_{1}>0$ is the characteristic radial length scale controlling the variation of the gravitational potential. The normalization $\Phi(r_{\mathrm{th}})=0$ is chosen without loss of generality, since it corresponds to a rescaling of the time coordinate and does not alter the local geometry. The present form is selected because it is analytic, symmetric with respect to the throat, and yields a vanishing first derivative at $r=r_{\mathrm{th}}$, thereby avoiding an artificial tidal singularity at the throat while still allowing nontrivial tidal behavior away from it.

Since the spacetime metric in Eq.~\eqref{3.12} is written in the $v$-parametrization with $r=\varrho(v)$, we distinguish throughout this section between derivatives with respect to $r$ and $v$ as follows: ${}^{\backprime}$ denotes $\frac{d}{dr}$ and ${}^{\prime}$ denotes $\frac{d}{dv}$. The chain rule then gives
\[
\Phi^{\prime}(v)
=
\Phi^{\backprime}(\varrho(v))\,\varrho^{\prime}(v),
\qquad
\Phi^{\prime\prime}(v)
=
\Phi^{\backprime\backprime}(\varrho(v))\,\bigl[\varrho^{\prime}(v)\bigr]^{2}
+
\Phi^{\backprime}(\varrho(v))\,\varrho^{\prime\prime}(v).
\]

Differentiating Eq.~\eqref{5.1} with respect to $r$ yields
\begin{equation}\label{5.2}
\Phi^{\backprime}(r)
=
\frac{\Phi_{0}}{\Phi_{1}}
\tanh\!\left(\frac{r-r_{\mathrm{th}}}{\Phi_{1}}\right),
\end{equation}
and a second differentiation gives
\begin{equation}\label{5.3}
\Phi^{\backprime\backprime}(r)
=
\frac{\Phi_{0}}{\Phi_{1}^{2}}
\operatorname{sech}^{2}\!\left(\frac{r-r_{\mathrm{th}}}{\Phi_{1}}\right).
\end{equation}
At the throat $r=r_{\mathrm{th}}$, one has
\[
\Phi^{\backprime}(r_{\mathrm{th}})=0,
\qquad
\Phi^{\backprime\backprime}(r_{\mathrm{th}})=\frac{\Phi_{0}}{\Phi_{1}^{2}}.
\]
Thus the chosen redshift function is regular at the throat and provides a finite tidal scale there, while still allowing nontrivial tidal gradients away from the throat.

For later use, it is convenient to note that Eq.~\eqref{5.1} satisfies the identity
\begin{equation}\label{sp1}
\Phi_{1}\,\sinh\!\left(\frac{r-r_{\mathrm{th}}}{\Phi_{1}}\right)\,
\Phi^{\backprime\backprime}(r)
-
\cosh\!\left(\frac{r-r_{\mathrm{th}}}{\Phi_{1}}\right)\,
\Phi^{\backprime}(r)
=0,
\end{equation}
which is simply a differential relation obeyed by the adopted smooth profile. This form makes explicit that the redshift function is controlled by a single length scale $\Phi_{1}$ and remains analytic throughout the domain of interest.

In the $v$-representation, the derivatives of the redshift function become
\begin{equation}\label{5.4}
\Phi^{\prime}(v)
=
\Phi^{\backprime}(\varrho(v))\,\varrho^{\prime}(v)
=
\frac{\Phi_{0}}{\Phi_{1}}
\tanh\!\left(\frac{\varrho(v)-r_{\mathrm{th}}}{\Phi_{1}}\right)
\varrho^{\prime}(v),
\end{equation}
and
\begin{equation}\label{5.5}
\Phi^{\prime\prime}(v)
=
\frac{\Phi_{0}}{\Phi_{1}^{2}}
\operatorname{sech}^{2}\!\left(\frac{\varrho(v)-r_{\mathrm{th}}}{\Phi_{1}}\right)
\bigl[\varrho^{\prime}(v)\bigr]^{2}
+
\frac{\Phi_{0}}{\Phi_{1}}
\tanh\!\left(\frac{\varrho(v)-r_{\mathrm{th}}}{\Phi_{1}}\right)
\varrho^{\prime\prime}(v).
\end{equation}
At the throat, $v=0$, one has $\varrho(0)=r_{\mathrm{th}}$ and $\varrho^{\prime}(0)=0$, so that
\[
\Phi^{\prime}(0)=0,
\qquad
\Phi^{\prime\prime}(0)=0.
\]
Hence the redshift profile introduces no singular tidal contribution at the throat in the $v$-parametrization, while still allowing nonzero tidal structure in a neighborhood of the throat.

Let indices $\hat{i}, \hat{j} \in \{\hat{v}, \hat{u}, \hat{\phi}\}$. The orthonormal-frame tidal accelerations for a static observer are governed by the Riemann tensor components \cite{thorne2000gravitation}. For a proper separation vector $\Delta \xi^{\hat{j}}$, the relative tidal acceleration satisfies
\[
\Delta \zeta^{\hat{i}}
=
-\mathcal{R}_{\hat{t}\hat{i}\hat{t}\hat{j}}\,
\Delta \xi^{\hat{j}}.
\]
In the present geometry, contracting the coordinate-basis Riemann tensor with the orthonormal tetrad
\[
e_{\hat{t}}=e^{-\Phi}\partial_{t},
\qquad
e_{\hat{v}}=\bigl(\varrho^{\prime}(v)^{2}+4a^{2}\bigr)^{-1/2}\partial_{v},
\qquad
e_{\hat{u}}=\frac{1}{\varrho}\partial_{u},
\]
which is unit-normalized with respect to the metric \eqref{3.12}, yields the relevant nonzero orthonormal components
\begin{equation}\label{5.6}
\begin{aligned}
\mathcal{R}_{\hat{t}\hat{v}\hat{t}\hat{v}}
&=
\frac{1}{\varrho^{\prime}(v)^{2}+4a^{2}}
\left[
\Phi^{\prime\prime}(v)
+
\bigl(\Phi^{\prime}(v)\bigr)^{2}
-
\frac{\varrho^{\prime}(v)\,\varrho^{\prime\prime}(v)}
{\varrho^{\prime}(v)^{2}+4a^{2}}
\,\Phi^{\prime}(v)
\right],
\\[4pt]
\mathcal{R}_{\hat{t}\hat{u}\hat{t}\hat{u}}
&=
\frac{\varrho^{\prime}(v)}
{\varrho^{\prime}(v)^{2}+4a^{2}}\,
\frac{\Phi^{\prime}(v)}{\varrho(v)}.
\end{aligned}
\end{equation}

The components $\mathcal{R}_{\hat{t}\hat{v}\hat{t}\hat{v}}$ and $\mathcal{R}_{\hat{t}\hat{u}\hat{t}\hat{u}}$ determine the radial and transverse tidal accelerations per unit separation. In geometric units ($G=c=1$), they have dimensions of inverse length squared. In SI units, the magnitude of the tidal acceleration experienced by a traveler separated by a proper distance $\Delta L$ is estimated by
\begin{equation}\label{5.7}
\Delta \zeta
=
c^{2}\,|\mathcal{R}|\,\Delta L,
\end{equation}
where $\mathcal{R}$ denotes one of the orthonormal components in Eq.~\eqref{5.6}. Because $\Phi^{\prime}(0)=0$, $\Phi^{\prime\prime}(0)=0$, and $\varrho^{\prime}(0)=0$, both the transverse and radial tidal components vanish identically at the throat. Thus the chosen redshift function does not introduce any divergent, or even nonzero, tidal spike at the throat. Its main physical role is to provide a controlled tidal sector that modifies the pressure distribution away from the throat and can therefore reduce the radial extent of exotic matter, although it does not eliminate the throat-level flare-out requirement itself.

For traversability, the tidal acceleration should remain below the terrestrial gravitational acceleration $g_{\oplus}\simeq 9.8~\mathrm{m/s^{2}}$ for a representative separation, say $\Delta L=2~\mathrm{m}$. This gives the estimate
\begin{equation}\label{5.8}
    |\mathcal{R}| \leq \frac{g_{\otimes}}{c^{2} \ \Delta L} \approx \frac{9.8}{(2.99792458 \ \times \ 10^{8})^{2} \ \times 2} \ \approx \ 5.45198527 \ \times \ 10^{-17} \ \ m^{-2}.
\end{equation}
Hence the redshift profile must vary slowly enough that the associated orthonormal curvature components remain below this order of magnitude in the traveler’s local frame.

In summary, the non-constant redshift function \eqref{5.1} is chosen because it is smooth, throat-symmetric, and controlled by a single tidal scale $\Phi_{1}$. It provides a finite and adjustable tidal sector, redistributes the radial pressure away from the throat, and may reduce the amount of exotic matter required in a neighborhood of the throat, while preserving the local flare-out structure imposed by the embedding geometry.

\section{\label{sec:level5}Field Equations and Energy Conditions}

Consider the coordinate system $x^{\mu}=(t,v,u,\phi)$. From the generalized super--hyperbolic metric \eqref{3.12}, the non-zero components of the metric tensor are
\begin{equation}\label{4.1}
    g_{\mu\nu}
    =
    \mathrm{diag}\!\left(
    -e^{2\Phi(v)},
    \ \varrho^{\prime}(v)^{2}+4a^{2},
    \ \varrho(v)^{2},
    \ \varrho(v)^{2}\sin^{2}u
    \right),
\end{equation}
so that the determinant is
\begin{equation}\label{4.2}
    \sqrt{-g}
    =
    e^{\Phi(v)}\,\varrho(v)^{2}\sin u\,
    \sqrt{\varrho^{\prime}(v)^{2}+4a^{2}}.
\end{equation}

The Christoffel symbols are defined by
\begin{equation}\label{4.3}
    \Gamma^{\mu}{}_{\nu\lambda}
    =
    \frac{1}{2}g^{\mu\sigma}
    \left(
    \partial_{\nu}g_{\sigma\lambda}
    +
    \partial_{\lambda}g_{\sigma\nu}
    -
    \partial_{\sigma}g_{\nu\lambda}
    \right),
\end{equation}
where $\partial_{\lambda}=\frac{\partial}{\partial x^{\lambda}}$. For the present diagonal, $v$-dependent metric, the non-zero Christoffel symbols are
\begin{equation}\label{4.4}
    \begin{aligned}
        \Gamma^{t}{}_{tv} &= \Phi^{\prime}(v),\\
        \Gamma^{v}{}_{tt} &= \frac{e^{2\Phi(v)}\,\Phi^{\prime}(v)}{\varrho^{\prime}(v)^{2}+4a^{2}},
        \qquad
        \Gamma^{v}{}_{vv} = \frac{\varrho^{\prime}(v)\,\varrho^{\prime\prime}(v)}{\varrho^{\prime}(v)^{2}+4a^{2}},\\
        \Gamma^{v}{}_{uu} &= -\frac{\varrho(v)\,\varrho^{\prime}(v)}{\varrho^{\prime}(v)^{2}+4a^{2}},
        \qquad
        \Gamma^{v}{}_{\phi\phi}
        = -\frac{\varrho(v)\,\varrho^{\prime}(v)}{\varrho^{\prime}(v)^{2}+4a^{2}}\sin^{2}u,\\
        \Gamma^{u}{}_{vu} &= \Gamma^{\phi}{}_{v\phi} = \frac{\varrho^{\prime}(v)}{\varrho(v)},
        \qquad
        \Gamma^{u}{}_{\phi\phi} = -\sin u\,\cos u,
        \qquad
        \Gamma^{\phi}{}_{u\phi} = \cot u.
    \end{aligned}
\end{equation}
All other components vanish by symmetry, since the metric is static and depends only on $v$.

By making use of the Riemann curvature tensor, 
\begin{equation}\label{4.5}
    R^{\gamma}_{\ \ \sigma \mu \nu} = \partial_{\mu}  \Gamma^{\gamma}_{\ \nu\sigma} - \partial_{\nu}\Gamma^{\gamma}_{\ \mu\sigma} + \Gamma^{\gamma}_{\ \mu \lambda} \Gamma^{\lambda}_{\ \nu \sigma} - \Gamma^{\gamma}_{\ \nu\lambda} \Gamma^{\lambda}_{\ \mu\sigma},
\end{equation}
and noting that all metric functions depend only on $v$, direct evaluation from Eq.~\eqref{4.4} gives the independent non-zero components as
\begin{equation}\label{4.6}
    \begin{aligned}
        R^{t}{}_{vtv}
        &=
        -\Phi^{\prime\prime}
        -
        (\Phi^{\prime})^{2}
        +
        \frac{\varrho^{\prime}\,\varrho^{\prime\prime}}{\varrho^{\prime 2}+4a^{2}}\,\Phi^{\prime},
        \\
        R^{t}{}_{utu}
        &=
        -\frac{\varrho\,\varrho^{\prime}}{\varrho^{\prime 2}+4a^{2}}\,\Phi^{\prime},
        \\
        R^{v}{}_{uvu}
        &=
        -\frac{4a^{2}\,\varrho\,\varrho^{\prime\prime}}{(\varrho^{\prime 2}+4a^{2})^{2}},
        \\
        R^{u}{}_{\phi u\phi}
        &=
        \sin^{2}u\left(\frac{4a^{2}}{\varrho^{\prime 2}+4a^{2}}\right).
    \end{aligned}
\end{equation}
The angular block carries the spherical curvature through the factor $\frac{4a^{2}}{{(\varrho^{\prime}}^{2} + 4a^{2})}$.

Contracting the Riemann tensor gives the Ricci tensor $R_{\mu\nu}=R^{\lambda}{}_{\mu\lambda\nu}$, whose non-zero components are
\begin{equation}\label{4.7}
    \begin{aligned}
        R_{tt}
        &=
        \frac{e^{2\Phi}}{\varrho^{\prime 2}+4a^{2}}
        \left[
        \Phi^{\prime\prime}
        +(\Phi^{\prime})^{2}
        +\frac{2\Phi^{\prime}\varrho^{\prime}}{\varrho}
        -\frac{\Phi^{\prime}\varrho^{\prime}\varrho^{\prime\prime}}{\varrho^{\prime 2}+4a^{2}}
        \right],
        \\
        R_{vv}
        &=
        -\Phi^{\prime\prime}-(\Phi^{\prime})^{2}
        +
        \frac{\Phi^{\prime}\varrho^{\prime}\varrho^{\prime\prime}}{\varrho^{\prime 2}+4a^{2}}
        -
        \frac{8a^{2}\varrho^{\prime\prime}}{\varrho(\varrho^{\prime 2}+4a^{2})},
        \\
        R_{uu}
        &=
        \frac{4a^{2}-\varrho\,\Phi^{\prime}\varrho^{\prime}}{\varrho^{\prime 2}+4a^{2}}
        -
        \frac{4a^{2}\varrho\,\varrho^{\prime\prime}}{(\varrho^{\prime 2}+4a^{2})^{2}},
        \\
        R_{\phi\phi}
        &=
        R_{uu}\sin^{2}u.
    \end{aligned}
\end{equation}
The Ricci scalar, $R= g^{\mu\nu}R_{\mu\nu}$, becomes
\begin{equation}\label{4.8}
    R(v)
    =
    -\frac{4}{\varrho^{\prime 2}+4a^{2}}
    \left[
    \frac{1}{2}\Bigl(\Phi^{\prime\prime}+(\Phi^{\prime})^{2}\Bigr)
    +
    \frac{\Phi^{\prime}\varrho^{\prime}}{\varrho}
    -
    \frac{\Phi^{\prime}\varrho^{\prime}\varrho^{\prime\prime}}{2(\varrho^{\prime 2}+4a^{2})}
    +
    \frac{4a^{2}\varrho^{\prime\prime}}{\varrho(\varrho^{\prime 2}+4a^{2})}
    -
    \frac{2a^{2}}{\varrho^{2}}
    \right].
\end{equation}

The EFEs in geometrical units are represented by \cite{einstein1915feldgleichungen, chandrasekhar1998mathematical, thorne2014black, hawking2023large},
\begin{equation}\label{4.9}
    G_{\mu\nu} = R_{\mu\nu} - \frac{1}{2} g_{\mu\nu} R
 = 8\pi T_{\mu\nu},
\end{equation}
where $G_{\mu\nu}$ denotes the Einstein tensor and for an anisotropic perfect fluid model containing the matter $(\mathrm{m})$, $T_{\mu\nu}$ represents the energy-momentum tensor in a diagonal form as
\begin{equation}\label{4.10}
    T^{\mu}_{\ \ \nu} = \text{diag} [-\rho_{\mathrm{m}},\ p_{r},\ p_{t},\ p_{t}].
\end{equation}
Here, $\rho_{\mathrm{m}}$ denotes the matter energy density in the system, $p_{r}$ and $p_{t}$ denote the radial and tangential pressure of the embedded surface, respectively. Using Eqs.~\eqref{4.9} and \eqref{4.10}, the field equations become
\begin{equation}\label{4.11}
    8\pi \rho_{\mathrm{m}}
    =
    \frac{4a^{2}}{(\varrho^{\prime 2}+4a^{2})^{2}}
    \cdot
    \frac{\bigl(\varrho^{\prime 2}+4a^{2}-2\varrho\,\varrho^{\prime\prime}\bigr)}{\varrho^{2}},
\end{equation}
\begin{equation}\label{4.12}
    8\pi p_{r}(v)
    =
    -\frac{4a^{2}}{\varrho^{2}(\varrho^{\prime 2}+4a^{2})}
    +
    \frac{2\Phi^{\prime}(v)\varrho^{\prime}(v)}{\varrho(\varrho^{\prime 2}+4a^{2})},
\end{equation}
\begin{equation}\label{4.13}
    8\pi p_{t}(v)
    =
    \frac{1}{\varrho^{\prime 2}+4a^{2}}
    \left[
    \Phi^{\prime\prime}
    +
    (\Phi^{\prime})^{2}
    +
    \frac{\Phi^{\prime}\varrho^{\prime}}{\varrho}
    -
    \frac{\Phi^{\prime}\varrho^{\prime}\varrho^{\prime\prime}}{\varrho^{\prime 2}+4a^{2}}
    \right]
    +
    \frac{4a^{2}\varrho^{\prime\prime}}{\varrho(\varrho^{\prime 2}+4a^{2})^{2}}.
\end{equation}

Equation \eqref{4.13} is best interpreted as the sum of the same three tidal contributions as before, together with a fourth term of different geometric origin. The first term in the square bracket,
\[
\Phi^{\prime\prime},
\]
measures the local curvature of the redshift profile, while the second,
\[
(\Phi^{\prime})^{2},
\]
represents the quadratic tidal contribution. The mixed term
\[
\frac{\Phi^{\prime}\varrho^{\prime}}{\varrho}
\]
couples the tidal gradient to the radial growth of the embedding surface. The fourth term inside the bracket,
\[
-\frac{\Phi^{\prime}\varrho^{\prime}\varrho^{\prime\prime}}{\varrho^{\prime 2}+4a^{2}},
\]
is a second-order tidal--curvature cross term, coupling the redshift gradient to the embedding's own curvature. The final term outside the square bracket,
\[
\frac{4a^{2}\varrho^{\prime\prime}}{\varrho(\varrho^{\prime 2}+4a^{2})^{2}},
\]
is the purely geometric contribution of the throat curvature to the transverse pressure. Thus $p_{t}(v)$ is determined by a balance between tidal effects and the embedding curvature, with no separate contribution from the angular sector itself once $k=+1$ is adopted.

At the throat, $v=0$, one has
\[
\varrho(0)=r_{\mathrm{th}}=\varrho_{0},
\qquad
\varrho^{\prime}(0)=0,
\qquad
\varrho^{\prime\prime}(0)=2\kappa,
\]
and therefore the redshift sector is regular there. In the present parametrization,
\[
\Phi^{\prime}(0)=0,
\qquad
\Phi^{\prime\prime}(0)=0,
\]
so the throat values are controlled entirely by the geometric terms.

Energy density at the throat arises entirely from geometry. There is no kinetic component (i.e, the wormhole is static), and redshift is irrelevant at the throat. Evaluating Eq.~\eqref{4.11} at the throat gives
\begin{equation}\label{gpsm1}
  8\pi\rho_{m}(0)=\frac{a^{2}-\varrho_{0}\kappa}{a^{2}\varrho_{0}^{2}},
\end{equation}
so that the numerator is a competition between $a^{2}$, set by the Mittag--Leffler amplitude, and $\varrho_{0}\kappa$, set by the throat radius together with the local curvature parameter of Sec.~\ref{subsec:lavel2.2}. If $a^{2}>\varrho_{0}\kappa$, the throat carries positive energy density; if $a^{2}<\varrho_{0}\kappa$, the density is negative, analogous to a Casimir-type vacuum. Unlike the original construction, this sign is controlled by $\kappa$ and $a$ rather than by the deformation parameter $\alpha$, which no longer enters the throat value of $\rho_{m}$ at all.

The radial pressure at the throat remains independent of both $\alpha$ and $\kappa$, which is a pure geometry constraint from the fact that $p_{r}(0) = \frac{-1}{8\pi \varrho_{0}^{2}}$, for $b(r_{th}) = r_{th}=\varrho_{0}$. The negative sign shows that the radial stress satisfies $p_{r}(0)<0$, making it radial tension. The physicality implies that the matter is always pulling inward, which is counteracting the throat's tendency to collapse. This is required for wormhole stability from the flaring-out condition, \emph{the matter must be radially tensile if and only if the rate of change of the shape function with respect to the throat radius is always less than 1 ($b^{\backprime}(r_{th})<1$)}.

In particular,
\begin{equation}\label{sp2}
    8\pi p_{t}(0)
    =
    \frac{\kappa}{2a^{2}\varrho_{0}}.
\end{equation}
This expression makes the physical interpretation transparent: it is unconditionally positive, since $\kappa,a,\varrho_{0}>0$ are all required by the admissibility axioms of Sec.~\ref{subsec:lavel2.2}. Unlike the original construction, the sign of $p_{t}(0)$ is no longer a competition between geometric scale and fractional deformation, since the Mittag--Leffler deformation contributes nothing to any throat quantity for $\alpha>1$ (Eq.~\eqref{2.7}); the transverse pressure at the throat is fixed entirely by the local curvature $\kappa$. Hence, Eq.~\eqref{sp2} shows that angular stability at the throat is governed by $\kappa$ and $a$, not by the higher-order deformation parameter $\alpha$.

The null energy condition (NEC) requires
\[
\rho_{\mathrm{m}}+p_{r}\geq 0.
\]
Using $\varrho^{\prime}(0)=0$, one obtains at the throat
\begin{equation}\label{4.14}
    8\pi\,[\rho_{\mathrm{m}}+p_{r}]_{\mathrm{th}}
    =
    \frac{b^{\backprime}(r_{\mathrm{th}})-1}{r_{\mathrm{th}}^{2}}
    =
    -\frac{\varrho^{\prime\prime}(0)}{2a^{2}r_{\mathrm{th}}}
    =
    -\frac{\kappa}{a^{2}\varrho_{0}}
    <0.
\end{equation}
Because $\kappa>0$ is required by Axiom 3 (Sec.~\ref{subsec:lavel2.2}) for every admissible embedding, the NEC is violated at the throat unconditionally, rather than only for $\alpha>\frac{1}{2}$ as in the original construction. This violation is not a failure of the construction; rather, it is the standard flare--out requirement expressed in matter form. Moreover, the magnitude of the violation is now controlled by $\kappa$ and $a$: a smaller local curvature $\kappa$, or a larger Mittag--Leffler amplitude $a$, reduces the required exoticity, whereas the deformation parameter $\alpha$ no longer enters this throat-level result at all.

\begin{figure}
\centering
\begin{tikzpicture}
\begin{axis}[
    width=0.92\linewidth, height=6.5cm,
    xlabel={$v$}, ylabel={$8\pi\times(\text{density, pressures})$},
    xmin=-2.5, xmax=2.5, ymin=-1.1, ymax=0.9,
    axis lines=left,
    legend style={at={(1.02,1)}, anchor=north west, font=\footnotesize, draw=none},
    label style={font=\small}, tick label style={font=\footnotesize},
    every axis plot/.append style={line width=1.15pt},
]
\addplot[gray, dotted, line width=0.6pt, domain=-2.5:2.5, forget plot] {0};
\addplot[black!60!white, dashdotted, line width=0.9pt] coordinates {(-2.5,0.1345) (-2.45,0.1392) (-2.4,0.1439) (-2.35,0.1486) (-2.3,0.1532) (-2.25,0.1577) (-2.2,0.162) (-2.15,0.166) (-2.1,0.1697) (-2.05,0.173) (-2.0,0.1758) (-1.95,0.1778) (-1.9,0.179) (-1.85,0.1792) (-1.8,0.1781) (-1.75,0.1755) (-1.7,0.1712) (-1.65,0.1649) (-1.6,0.1562) (-1.55,0.1449) (-1.5,0.1308) (-1.45,0.1136) (-1.4,0.09305) (-1.35,0.06913) (-1.3,0.04182) (-1.25,0.01118) (-1.2,-0.02258) (-1.15,-0.05916) (-1.1,-0.09814) (-1.05,-0.139) (-1.0,-0.1812) (-0.95,-0.2239) (-0.9,-0.2667) (-0.85,-0.3086) (-0.8,-0.3489) (-0.75,-0.3869) (-0.7,-0.422) (-0.65,-0.4533) (-0.6,-0.4804) (-0.55,-0.5027) (-0.5,-0.5197) (-0.45,-0.5311) (-0.4,-0.5367) (-0.35,-0.5362) (-0.3,-0.5297) (-0.25,-0.5169) (-0.2,-0.4977) (-0.15,-0.4717) (-0.1,-0.4377) (-0.05,-0.3922) (0.0,-0.3) (0.05,-0.3922) (0.1,-0.4377) (0.15,-0.4717) (0.2,-0.4977) (0.25,-0.5169) (0.3,-0.5297) (0.35,-0.5362) (0.4,-0.5367) (0.45,-0.5311) (0.5,-0.5197) (0.55,-0.5027) (0.6,-0.4804) (0.65,-0.4533) (0.7,-0.422) (0.75,-0.3869) (0.8,-0.3489) (0.85,-0.3086) (0.9,-0.2667) (0.95,-0.2239) (1.0,-0.1812) (1.05,-0.139) (1.1,-0.09814) (1.15,-0.05916) (1.2,-0.02258) (1.25,0.01118) (1.3,0.04182) (1.35,0.06913) (1.4,0.09305) (1.45,0.1136) (1.5,0.1308) (1.55,0.1449) (1.6,0.1562) (1.65,0.1649) (1.7,0.1712) (1.75,0.1755) (1.8,0.1781) (1.85,0.1792) (1.9,0.179) (1.95,0.1778) (2.0,0.1758) (2.05,0.173) (2.1,0.1697) (2.15,0.166) (2.2,0.162) (2.25,0.1577) (2.3,0.1532) (2.35,0.1486) (2.4,0.1439) (2.45,0.1392) (2.5,0.1345)};
\addlegendentry{NEC ($\rho_m+p_r$)}
\addplot[teal, solid] coordinates {(-2.5,-0.00157) (-2.45,-0.001821) (-2.4,-0.002108) (-2.35,-0.002438) (-2.3,-0.002814) (-2.25,-0.003243) (-2.2,-0.003729) (-2.15,-0.00428) (-2.1,-0.004901) (-2.05,-0.005599) (-2.0,-0.006378) (-1.95,-0.007245) (-1.9,-0.008203) (-1.85,-0.009254) (-1.8,-0.0104) (-1.75,-0.01163) (-1.7,-0.01293) (-1.65,-0.0143) (-1.6,-0.01571) (-1.55,-0.01711) (-1.5,-0.01846) (-1.45,-0.0197) (-1.4,-0.02073) (-1.35,-0.02145) (-1.3,-0.02173) (-1.25,-0.02139) (-1.2,-0.02026) (-1.15,-0.01809) (-1.1,-0.01465) (-1.05,-0.009647) (-1.0,-0.00279) (-0.95,0.006225) (-0.9,0.0177) (-0.85,0.03191) (-0.8,0.0491) (-0.75,0.06944) (-0.7,0.09302) (-0.65,0.1199) (-0.6,0.1499) (-0.55,0.1828) (-0.5,0.2183) (-0.45,0.2561) (-0.4,0.2957) (-0.35,0.3365) (-0.3,0.3783) (-0.25,0.4206) (-0.2,0.4636) (-0.15,0.5074) (-0.1,0.5535) (-0.05,0.6058) (0.0,0.7) (0.05,0.6058) (0.1,0.5535) (0.15,0.5074) (0.2,0.4636) (0.25,0.4206) (0.3,0.3783) (0.35,0.3365) (0.4,0.2957) (0.45,0.2561) (0.5,0.2183) (0.55,0.1828) (0.6,0.1499) (0.65,0.1199) (0.7,0.09302) (0.75,0.06944) (0.8,0.0491) (0.85,0.03191) (0.9,0.0177) (0.95,0.006225) (1.0,-0.00279) (1.05,-0.009647) (1.1,-0.01465) (1.15,-0.01809) (1.2,-0.02026) (1.25,-0.02139) (1.3,-0.02173) (1.35,-0.02145) (1.4,-0.02073) (1.45,-0.0197) (1.5,-0.01846) (1.55,-0.01711) (1.6,-0.01571) (1.65,-0.0143) (1.7,-0.01293) (1.75,-0.01163) (1.8,-0.0104) (1.85,-0.009254) (1.9,-0.008203) (1.95,-0.007245) (2.0,-0.006378) (2.05,-0.005599) (2.1,-0.004901) (2.15,-0.00428) (2.2,-0.003729) (2.25,-0.003243) (2.3,-0.002814) (2.35,-0.002438) (2.4,-0.002108) (2.45,-0.001821) (2.5,-0.00157)};
\addlegendentry{$8\pi\rho_m$}
\addplot[purple, solid] coordinates {(-2.5,0.1361) (-2.45,0.1411) (-2.4,0.1461) (-2.35,0.151) (-2.3,0.156) (-2.25,0.1609) (-2.2,0.1657) (-2.15,0.1703) (-2.1,0.1746) (-2.05,0.1786) (-2.0,0.1821) (-1.95,0.1851) (-1.9,0.1872) (-1.85,0.1884) (-1.8,0.1885) (-1.75,0.1872) (-1.7,0.1842) (-1.65,0.1792) (-1.6,0.1719) (-1.55,0.1621) (-1.5,0.1493) (-1.45,0.1333) (-1.4,0.1138) (-1.35,0.09059) (-1.3,0.06355) (-1.25,0.03257) (-1.2,-0.002326) (-1.15,-0.04107) (-1.1,-0.08349) (-1.05,-0.1294) (-1.0,-0.1784) (-0.95,-0.2302) (-0.9,-0.2844) (-0.85,-0.3405) (-0.8,-0.398) (-0.75,-0.4564) (-0.7,-0.515) (-0.65,-0.5732) (-0.6,-0.6303) (-0.55,-0.6855) (-0.5,-0.738) (-0.45,-0.7872) (-0.4,-0.8323) (-0.35,-0.8728) (-0.3,-0.9079) (-0.25,-0.9375) (-0.2,-0.9612) (-0.15,-0.9791) (-0.1,-0.9912) (-0.05,-0.998) (0.0,-1.0) (0.05,-0.998) (0.1,-0.9912) (0.15,-0.9791) (0.2,-0.9612) (0.25,-0.9375) (0.3,-0.9079) (0.35,-0.8728) (0.4,-0.8323) (0.45,-0.7872) (0.5,-0.738) (0.55,-0.6855) (0.6,-0.6303) (0.65,-0.5732) (0.7,-0.515) (0.75,-0.4564) (0.8,-0.398) (0.85,-0.3405) (0.9,-0.2844) (0.95,-0.2302) (1.0,-0.1784) (1.05,-0.1294) (1.1,-0.08349) (1.15,-0.04107) (1.2,-0.002326) (1.25,0.03257) (1.3,0.06355) (1.35,0.09059) (1.4,0.1138) (1.45,0.1333) (1.5,0.1493) (1.55,0.1621) (1.6,0.1719) (1.65,0.1792) (1.7,0.1842) (1.75,0.1872) (1.8,0.1885) (1.85,0.1884) (1.9,0.1872) (1.95,0.1851) (2.0,0.1821) (2.05,0.1786) (2.1,0.1746) (2.15,0.1703) (2.2,0.1657) (2.25,0.1609) (2.3,0.156) (2.35,0.151) (2.4,0.1461) (2.45,0.1411) (2.5,0.1361)};
\addlegendentry{$8\pi p_r$}
\addplot[orange!90!black, solid] coordinates {(-2.5,0.3033) (-2.45,0.3051) (-2.4,0.307) (-2.35,0.309) (-2.3,0.311) (-2.25,0.3132) (-2.2,0.3154) (-2.15,0.3178) (-2.1,0.3203) (-2.05,0.3231) (-2.0,0.326) (-1.95,0.3292) (-1.9,0.3327) (-1.85,0.3365) (-1.8,0.3407) (-1.75,0.3452) (-1.7,0.35) (-1.65,0.3552) (-1.6,0.3606) (-1.55,0.3661) (-1.5,0.3716) (-1.45,0.3769) (-1.4,0.3817) (-1.35,0.3859) (-1.3,0.3891) (-1.25,0.3912) (-1.2,0.392) (-1.15,0.3914) (-1.1,0.3895) (-1.05,0.3864) (-1.0,0.3821) (-0.95,0.3769) (-0.9,0.3711) (-0.85,0.3648) (-0.8,0.3582) (-0.75,0.3516) (-0.7,0.3449) (-0.65,0.3382) (-0.6,0.3315) (-0.55,0.3246) (-0.5,0.3173) (-0.45,0.3094) (-0.4,0.3009) (-0.35,0.2914) (-0.3,0.2806) (-0.25,0.2685) (-0.2,0.2547) (-0.15,0.2388) (-0.1,0.22) (-0.05,0.1963) (0.0,0.15) (0.05,0.1963) (0.1,0.22) (0.15,0.2388) (0.2,0.2547) (0.25,0.2685) (0.3,0.2806) (0.35,0.2914) (0.4,0.3009) (0.45,0.3094) (0.5,0.3173) (0.55,0.3246) (0.6,0.3315) (0.65,0.3382) (0.7,0.3449) (0.75,0.3516) (0.8,0.3582) (0.85,0.3648) (0.9,0.3711) (0.95,0.3769) (1.0,0.3821) (1.05,0.3864) (1.1,0.3895) (1.15,0.3914) (1.2,0.392) (1.25,0.3912) (1.3,0.3891) (1.35,0.3859) (1.4,0.3817) (1.45,0.3769) (1.5,0.3716) (1.55,0.3661) (1.6,0.3606) (1.65,0.3552) (1.7,0.35) (1.75,0.3452) (1.8,0.3407) (1.85,0.3365) (1.9,0.3327) (1.95,0.3292) (2.0,0.326) (2.05,0.3231) (2.1,0.3203) (2.15,0.3178) (2.2,0.3154) (2.25,0.3132) (2.3,0.311) (2.35,0.309) (2.4,0.307) (2.45,0.3051) (2.5,0.3033)};
\addlegendentry{$8\pi p_t$}
\end{axis}
\end{tikzpicture}
\caption{\label{fig:energy}The matter density $8\pi\rho_m(v)$, radial pressure $8\pi p_r(v)$, transverse pressure $8\pi p_t(v)$, and the null-energy-condition combination $8\pi(\rho_m+p_r)$, for $\varrho_0=a=1$, $\kappa=0.3$, $\alpha=1.3$, $\Phi_0=0.5$, $\Phi_1=1$. The throat carries positive energy density while the NEC is violated; the violation is confined to a finite neighborhood of the throat, becoming satisfied for $|v|\gtrsim1.2$.}
\end{figure}
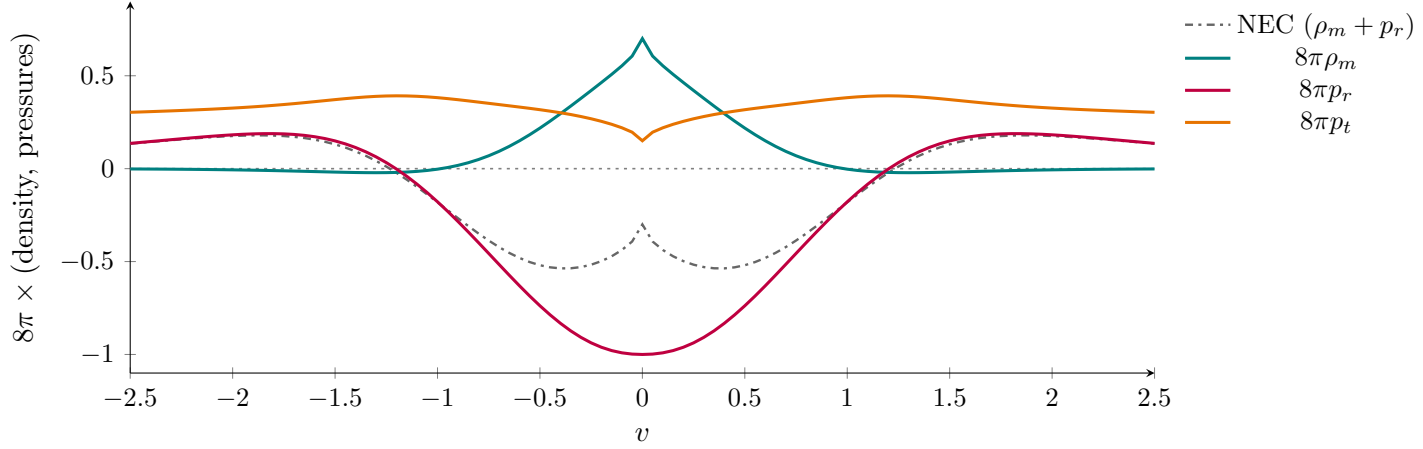

The weak energy condition (WEC) requires $\rho_{\mathrm{m}}\geq 0$ together with the NEC inequalities. At the throat, this reduces to the explicit condition $\kappa\leq a^{2}/\varrho_{0}$, obtained directly from the sign of $8\pi\rho_{m}(0)$ above; away from the throat, positivity of $\rho_{\mathrm{m}}$ from Eq.~\eqref{4.11} should be assessed for each admissible choice of $\alpha$, $a$, $\kappa$, $\varrho_{0}$, and $\Phi(v)$. Likewise, the strong energy condition (SEC),
\[
\rho_{\mathrm{m}}+p_{r}+2p_{t}\geq 0,
\]
is generally violated near the throat but may become less restrictive away from the throat if the tidal derivatives $\Phi^{\prime}$ and $\Phi^{\prime\prime}$ decay sufficiently rapidly and the embedding terms weaken in the asymptotic region. Thus, the non-constant redshift function does not remove the throat-level NEC violation implied by the flare--out condition, but it can redistribute the pressure profile and, for suitable parameter choices, reduce the radial extent of the exotic region.

The tidal interpretation is therefore the following: the redshift function \eqref{5.1} is chosen because it is smooth, symmetric about the throat, and controlled by a single length scale $\Phi_{1}$. It provides a finite tidal sector without introducing a divergence at $v=0$, and it allows the matter distribution to be adjusted in the neighborhood of the throat while preserving the geometric wormhole structure generated by the embedding.

\section{\label{sec:level6}Thin-Shell Formalism and Stability Analysis}

This section treats the thin shell as a \emph{junction surface} between the interior super--hyperbolic geometry and an exterior vacuum region. It should therefore be read as a matching construction rather than as a proof of global smoothness of the bulk throat. In particular, the thin-shell description is conceptually distinct from the regularity analysis carried out earlier for the embedding geometry. The purpose here is to obtain the surface stresses and the linearized radial stability criterion associated with the junction hypersurface.

Throughout this section, the interior spacetime is taken in the pure $v$-form,
\begin{equation}\label{6.1}
ds^{2}_{\text{(i)}}
=
-e^{2\Phi(v)}dt^{2}
+
\bigl[\varrho^{\prime}(v)^{2}+4a^{2}\bigr]\,dv^{2}
+
\varrho(v)^{2}\bigl(du^{2}+\sin^{2}u\,d\phi^{2}\bigr),
\end{equation}
with
\[
r\equiv \varrho(v)=\varrho_{0}+\kappa v^{2}+a\,\mathfrak{S}C_{\alpha}(v).
\] 
For the junction calculation, it is convenient to parametrize the shell by its areal radius
\[
r=f(\tau),
\]
where $\tau$ is the proper time measured on the shell. The embedding functions are
\[
t=t(\tau),\qquad v=v(\tau),\qquad u=\theta,\qquad \phi=\phi.
\]
Since $f(\tau)=\varrho(v(\tau))$, one has
\begin{equation}\label{6.2}
\dot f=\varrho^{\prime}(v)\,\dot v,
\end{equation}
where the dot denotes differentiation with respect to $\tau$. The induced metric on the hypersurface $\Sigma: r = f(\tau)$ is then
\begin{equation}\label{6.3}
ds^{2}_{\Sigma}
=
-d\tau^{2}
+
f(\tau)^{2}\bigl(d\theta^{2}+\sin^{2}\theta\,d\phi^{2}\bigr),
\end{equation}
which is the metric of a spherical timelike shell of radius $f(\tau)$. 

Consistency with the interior geometry requires the normalization condition
\begin{equation}\label{6.4}
-e^{2\Phi(v)}\dot t^{2}
+
\bigl[\varrho^{\prime}(v)^{2}+4a^{2}\bigr]\dot v^{2}
=
-1.
\end{equation}
Equivalently, if the interior is written in curvature coordinates along $\Sigma$, one may express the same condition in the standard form
\[
-A_{\mathrm{i}}(f)\dot t^{2}+\frac{\dot f^{2}}{B_{\mathrm{i}}(f)}=-1,
\]
with $A_{\mathrm{i}}(r)=e^{2\Phi(r)}$ and $B_{\mathrm{i}}(r)=1-b_{\mathrm{i}}(r)/r$.

The unit normal to $\Sigma$ is chosen outward on each side in the standard Israel--Lanczos convention \cite{israel1966singular}. The extrinsic curvature is defined by
\begin{equation}\label{6.5}
K_{ij}=-n_{\mu;\nu}\,e^{\mu}{}_{i}\,e^{\nu}{}_{j},
\end{equation}
where $e^{\mu}{}_{i}=\partial x^{\mu}/\partial\xi^{i}$ are the tangent basis vectors on $\Sigma$. For a diagonal and symmetry-adapted junction, only two independent components are required, namely $K^{\tau}{}_{\tau}$ and $K^{\theta}{}_{\theta}=K^{\phi}{}_{\phi}$. The interior components are
\begin{equation}\label{6.6}
{}^{(\mathrm{i})}K^{\theta}{}_{\theta}
=
\frac{1}{f}\sqrt{B_{\mathrm{i}}(f)+\dot f^{2}},
\end{equation}
and
\begin{equation}\label{6.7}
{}^{(\mathrm{i})}K^{\tau}{}_{\tau}
=
\frac{A_{\mathrm{i}}^{\prime}(f)+2\ddot f}
{2A_{\mathrm{i}}(f)\sqrt{B_{\mathrm{i}}(f)+\dot f^{2}}}.
\end{equation}
Here
\begin{equation}\label{6.8}
A_{\mathrm{i}}(r)=e^{2\Phi(r)},
\qquad
A_{\mathrm{i}}^{\backprime}(r)=2\Phi^{\backprime}(r)e^{2\Phi(r)},
\end{equation}
and
\begin{equation}\label{6.9}
B_{\mathrm{i}}(r)=1-\frac{b_{\mathrm{i}}(r)}{r},
\qquad
b_{\mathrm{i}}(r)
=
\frac{4a_{\mathrm{emb}}^{2}r}{\varrho^{\prime}(v(r))^{2}+4a_{\mathrm{emb}}^{2}},
\end{equation}
where $a_{\mathrm{emb}}$ is the embedding scale appearing in the interior geometry. The dependence of $B_{\mathrm{i}}$ on $\kappa$ and $\alpha$ is inherited through $\varrho(v)$ and its derivatives (Eqs.~\eqref{gps2}, \eqref{gps3}). The precise form of $B_{\mathrm{i}}^{\backprime}$ and $B_{\mathrm{i}}^{\backprime\backprime}$ is not needed explicitly here; only their existence and their $(\kappa,\alpha)$-dependence enter the stability analysis below.

For the exterior region, we adopt a vacuum spacetime of Schwarzschild type,
\begin{equation}\label{6.10}
ds^{2}_{\text{(e)}}
=
-A_{\mathrm{e}}(r)\,dt^{2}
+
\frac{dr^{2}}{B_{\mathrm{e}}(r)}
+
r^{2}\bigl(d\theta^{2}+\sin^{2}\theta\,d\phi^{2}\bigr),
\end{equation}
with
\begin{equation}\label{6.11}
A_{\mathrm{e}}(r)=B_{\mathrm{e}}(r)=1-\frac{2M}{r}.
\end{equation}
The corresponding exterior extrinsic curvature components are
\begin{equation}\label{6.12}
{}^{(\mathrm{e})}K^{\theta}{}_{\theta}
=
\frac{1}{f}\sqrt{B_{\mathrm{e}}(f)+\dot f^{2}},
\qquad
{}^{(\mathrm{e})}K^{\tau}{}_{\tau}
=
\frac{A_{\mathrm{e}}^{\prime}(f)+2\ddot f}
{2A_{\mathrm{e}}(f)\sqrt{B_{\mathrm{e}}(f)+\dot f^{2}}},
\end{equation}
where
\begin{equation}\label{6.13}
A_{\mathrm{e}}^{\prime}(f)=\frac{2M}{f^{2}}.
\end{equation}

The jump in any quantity across the shell is defined by
\[
[X]=X^{(\mathrm{e})}-X^{(\mathrm{i})}.
\]
Using the standard Israel--Lanczos junction formula \cite{israel1966singular},
\begin{equation}\label{6.14}
S^{i}{}_{j}
=
-\frac{1}{8\pi}\Bigl([K^{i}{}_{j}]-\delta^{i}{}_{j}[K]\Bigr),
\end{equation}
where $\delta^{i}{}_{j}$ is the Kronecker delta on $\Sigma$, and
\begin{equation}\label{6.15}
[K]=h^{ij}[K_{ij}]
=
[K^{\tau}{}_{\tau}]+2[K^{\theta}{}_{\theta}],
\end{equation}
with
\[
h_{ij}=\mathrm{diag}(-1,f^{2},f^{2}\sin^{2}\theta),
\]
the surface stress-energy tensor takes the diagonal form
\begin{equation}\label{6.16}
S^{i}{}_{j}
=
\mathrm{diag}(-\sigma,\mathcal P,\mathcal P).
\end{equation}
Thus $\sigma$ is the surface energy density and $\mathcal P$ is the surface pressure on the shell. The jumps in the extrinsic curvature are
\begin{equation}\label{6.17}
\begin{aligned}
[K^{\theta }{}_{\theta}]
&=
\frac{1}{f}
\left(
\sqrt{B_{\mathrm{e}}(f)+\dot f^{2}}
-
\sqrt{B_{\mathrm{i}}(f)+\dot f^{2}}
\right),
\\[4pt]
[K^{\tau}{}_{\tau}]
&=
\frac{A_{\mathrm{e}}^{\prime}(f)+2\ddot f}
{2A_{\mathrm{e}}(f)\sqrt{B_{\mathrm{e}}(f)+\dot f^{2}}}
-
\frac{A_{\mathrm{i}}^{\prime}(f)+2\ddot f}
{2A_{\mathrm{i}}(f)\sqrt{B_{\mathrm{i}}(f)+\dot f^{2}}}.
\end{aligned}
\end{equation}
Substituting these expressions into Eq.~\eqref{6.15} gives
\begin{equation}\label{6.18}
[K]
=
\frac{A_{\mathrm{e}}^{\prime}(f)+2\ddot f}
{2A_{\mathrm{e}}(f)\sqrt{B_{\mathrm{e}}(f)+\dot f^{2}}}
-
\frac{A_{\mathrm{i}}^{\prime}(f)+2\ddot f}
{2A_{\mathrm{i}}(f)\sqrt{B_{\mathrm{i}}(f)+\dot f^{2}}}
+
\frac{2}{f}
\left(
\sqrt{B_{\mathrm{e}}(f)+\dot f^{2}}
-
\sqrt{B_{\mathrm{i}}(f)+\dot f^{2}}
\right).
\end{equation}
Consequently, the surface energy density and pressure are
\begin{equation}\label{6.19}
\sigma(f)
=
-\frac{1}{4\pi f}
\left(
\sqrt{B_{\mathrm{e}}(f)+\dot f^{2}}
-
\sqrt{B_{\mathrm{i}}(f)+\dot f^{2}}
\right),
\end{equation}
and
\begin{equation}\label{6.20}
\begin{aligned}
\mathcal P(f)
&=
\frac{1}{8\pi}
\left[
\frac{A_{\mathrm{e}}^{\prime}(f)+2\ddot f}
{2A_{\mathrm{e}}(f)\sqrt{B_{\mathrm{e}}(f)+\dot f^{2}}}
-
\frac{A_{\mathrm{i}}^{\prime}(f)+2\ddot f}
{2A_{\mathrm{i}}(f)\sqrt{B_{\mathrm{i}}(f)+\dot f^{2}}}
\right]
\\[4pt]
&\quad
+
\frac{1}{8\pi f}
\left(
\sqrt{B_{\mathrm{e}}(f)+\dot f^{2}}
-
\sqrt{B_{\mathrm{i}}(f)+\dot f^{2}}
\right).
\end{aligned}
\end{equation}

At a static junction $f=f_{0}$, one has $\dot f=\ddot f=0$, and the above expressions reduce to
\[
\sigma_{0}
=
-\frac{1}{4\pi f_{0}}
\left(
\sqrt{B_{\mathrm{e}}(f_{0})}
-
\sqrt{B_{\mathrm{i}}(f_{0})}
\right),
\]
\[
\mathcal P_{0}
=
\frac{1}{8\pi}
\left[
\frac{A_{\mathrm{e}}^{\prime}(f_{0})}{2A_{\mathrm{e}}(f_{0})\sqrt{B_{\mathrm{e}}(f_{0})}}
-
\frac{A_{\mathrm{i}}^{\prime}(f_{0})}{2A_{\mathrm{i}}(f_{0})\sqrt{B_{\mathrm{i}}(f_{0})}}
+
\frac{\sqrt{B_{\mathrm{e}}(f_{0})}-\sqrt{B_{\mathrm{i}}(f_{0})}}{f_{0}}
\right].
\]
The sign of $\sigma_{0}$ depends on the relative size of the two square roots. In the wormhole context, a negative value is typically associated with the exotic surface support required at the junction, although this sign should be interpreted together with the full stability analysis rather than in isolation. The pressure $\mathcal P_{0}$ contains both curvature and redshift contributions; in particular, a positive $A_{\mathrm{i}}^{\prime}$, equivalently $\Phi^{\prime}(f_{0})>0$, can increase the outward pressure at the shell, but it does not by itself guarantee stability.

Define the surface mass by
\begin{equation}\label{6.21}
m_{s}(f)=4\pi f^{2}\sigma(f).
\end{equation}
Using Eq.~\eqref{6.19}, this becomes
\begin{equation}\label{6.22}
m_{s}(f)
=
-f\left(
\sqrt{B_{\mathrm{e}}(f)+\dot f^{2}}
-
\sqrt{B_{\mathrm{i}}(f)+\dot f^{2}}
\right).
\end{equation}
Rearranging Eq.~\eqref{6.22} yields
\begin{equation}\label{6.23}
\sqrt{B_{\mathrm{i}}(f)+\dot f^{2}}
=
\frac{m_{s}}{2f}
+
\frac{f\bigl(B_{\mathrm{i}}(f)-B_{\mathrm{e}}(f)\bigr)}{2m_{s}},
\end{equation}
and squaring once more gives
\begin{equation}\label{6.24}
B_{\mathrm{i}}(f)+\dot f^{2}
=
\left[
\frac{m_{s}}{2f}
+
\frac{f\bigl(B_{\mathrm{i}}(f)-B_{\mathrm{e}}(f)\bigr)}{2m_{s}}
\right]^{2}.
\end{equation}
Thus the shell motion can be written in the standard potential form,
\begin{equation}\label{6.25}
\dot f^{2}+V(f)=0,
\end{equation}
with effective potential
\begin{equation}\label{6.26}
V(f)
=
\frac{B_{\mathrm{i}}(f)+B_{\mathrm{e}}(f)}{2}
-
\left(\frac{m_{s}}{2f}\right)^{2}
-
\left(
\frac{f\,\Delta B(f)}{2m_{s}}
\right)^{2},
\qquad
\Delta B(f):={B_{\mathrm{i}}(f)-B_{\mathrm{e}}(f)}.
\end{equation}
This is the canonical thin-shell potential; the shell is static at $f=f_{0}$ when
\[
V(f_{0})=0,
\qquad
V^{\prime}(f_{0})=0.
\]
Linearized stability around $f_{0}$ is determined by the sign of $V^{\prime\prime}(f_{0})$: a positive value yields oscillatory behavior about equilibrium, while a negative value indicates linear instability in the radial sector.

The conservation law on the shell may be written as
\begin{equation}\label{6.27}
\left(\frac{m_{s}}{f}\right)^{\prime}
=
-4\pi\bigl(\sigma+2\mathcal P\bigr).
\end{equation}
If the shell equation of state is linearized near equilibrium as
\begin{equation}\label{6.28}
\mathcal P-\mathcal P_{0}
=
\eta\,(\sigma-\sigma_{0}),
\qquad
\eta:=\left(\frac{d\mathcal P}{d\sigma}\right)_{f_{0}},
\end{equation}
then one obtains
\begin{equation}\label{6.29}
\left(\frac{m_{s}}{f}\right)^{\prime\prime}
=
\frac{8\pi}{f}\,(1+2\eta)\,(\sigma+\mathcal P),
\end{equation}
evaluated at the static configuration. After standard algebra, the second derivative of the potential at equilibrium can be written as
\begin{equation}\label{6.30}
\begin{aligned}
V^{\prime\prime}(f_{0})
&=
B_{\mathrm{i}}^{\prime\prime}(f_{0})
-
\frac{1}{2B_{\mathrm{i}}(f_{0})}
\left[
\Delta B^{\prime}(f_{0})
-
4\pi f_{0}\bigl(\sigma_{0}+2\mathcal P_{0}\bigr)
\right]^{2}
\\[4pt]
&\quad
-
4\pi(1+2\eta)\bigl(\sigma_{0}+\mathcal P_{0}\bigr)
-
\frac{\Delta B^{\prime\prime}(f_{0})}{2}.
\end{aligned}
\end{equation}
This expression provides a local stability criterion for the shell under radial perturbations. The dependence on the interior geometry enters through $B_{\mathrm{i}}$ and its derivatives, hence through both the local curvature parameter $\kappa$ and the deformation parameter $\alpha$ inherited from the embedding of Sec.~\ref{subsec:lavel2.2}. The dependence on the redshift function enters through $A_{\mathrm{i}}$ and $A_{\mathrm{i}}^{\prime}$ in the static stresses. However, the stability condition should be interpreted carefully: it establishes linearized stability only in a neighbourhood of $f_{0}$ and for the chosen matching data. It does not by itself imply global stability or global smoothness of the full wormhole spacetime.

The main physical message of the thin-shell analysis is therefore the following. The shell radius $f(\tau)$ evolves under an effective potential determined by the mismatch between the interior and exterior geometries. The redshift profile controls the surface pressure through $A_{\mathrm{i}}^{\prime}$, and the shell equation of state introduces the material response parameter $\eta$. For a junction radius $f_{0}$ within the deformation-onset scale of Sec.~\ref{subsec:lavel2.3} (i.e., $|v(f_{0})|\lesssim v_{*}$), the interior curvature entering $B_{\mathrm{i}}^{\prime\prime}$ is governed primarily by the local curvature parameter $\kappa$, with the Mittag--Leffler deformation contributing only at higher order; for $f_{0}$ well beyond this scale, the deformation parameter $\alpha$ instead governs the interior curvature, through the amplitude of the far-field growth in Eq.~\eqref{2.14}. In either regime, the resulting sign of $V^{\prime\prime}(f_{0})$ must be verified case by case for the specific parameter choice and junction radius adopted. The thin-shell formalism thus provides a complementary, local dynamical diagnostic for the junction surface, distinct from the bulk regularity analysis of the interior geometry.

\section{\label{sec:level7}Conclusion}

In the present work, we formulated and analyzed a generalized class of traversable wormhole geometries generated from a super--hyperbolic deformation of the classical catenoidal embedding governed by a Mittag--Leffler-type structure. The construction extends the standard minimal-surface approach by introducing a deformation parameter $\alpha>1$, together with an explicit local curvature parameter $\kappa$ securing the throat, through which the embedding profile, shape function, and curvature structure inherit nontrivial geometric modifications relative to the classical Morris--Thorne configuration. Unlike conventional zero--tidal-force wormholes, the present framework incorporates a non-constant redshift function, thereby allowing the spacetime to possess a finite and controllable tidal sector.

Starting from the generalized embedding geometry, we derived the corresponding wormhole spacetime both in embedding coordinates and in curvature coordinates. The associated metric coefficients, Christoffel symbols, Riemann tensor, Ricci tensor, Ricci scalar, and Einstein tensor were obtained explicitly within the adopted coordinate framework. Particular attention was given to the geometric interpretation of the throat structure, secured by an explicit local curvature parameter $\kappa$ rather than by the deformation parameter $\alpha$, and to the resulting flare--out condition, shown to hold unconditionally for every admissible embedding. The resulting analysis demonstrated that $\kappa$ governs the throat's local curvature, while $\alpha$ acts as a geometric regulator of the embedding away from the throat, controlling the radial onset of the Mittag--Leffler deformation and the amplitude of its growth in the far field, and continuously approaching the classical catenoidal limit as $\alpha\rightarrow1^{+}$.

The matter sector supporting the wormhole was investigated through the Einstein field equations for an anisotropic stress--energy tensor, derived consistently for the spherically symmetric angular sector throughout. Explicit expressions for the matter density, radial pressure, and transverse pressure were derived and analyzed together with the corresponding null, weak, and strong energy conditions. The analysis confirmed that the flare--out condition is associated with local NEC violation at the throat, now shown to hold unconditionally for any admissible embedding, as expected for traversable wormhole geometries. At the same time, the non-constant redshift function modifies the pressure distribution away from the throat and can reduce the radial extent of the exotic region for suitable parameter choices, although it does not remove the throat-level flare--out requirement itself.

The junction of the interior super--hyperbolic geometry to an exterior Schwarzschild spacetime, now consistently spherically symmetric on both sides, was then studied using the Israel--Lanczos thin-shell formalism. The corresponding surface stress-energy tensor, shell equation of motion, and effective potential were derived explicitly. Linearized radial stability was analyzed through the condition $V^{\prime\prime}(f_{0})>0$, revealing that the stability properties depend jointly on the local curvature parameter $\kappa$, the deformation parameter $\alpha$, the tidal redshift structure, the shell equation-of-state parameter $\eta$, and the throat radius, with $\kappa$ governing junction radii near the throat and $\alpha$ governing those well beyond the deformation-onset scale. Within the parameter ranges considered here, the resulting stability domain remains local and must be verified case by case for the chosen junction radius.

Overall, the present framework provides a geometrically motivated extension of catenoidal wormhole embeddings in which the Mittag--Leffler deformation and tidal structure interact nontrivially with the matter and stability sectors. While several aspects of the construction remain sensitive to the detailed behavior of the throat geometry and the adopted matching conditions, the analysis developed here establishes a consistent basis for exploring generalized embedding-induced wormhole structures within relativistic gravitation. Possible future directions include the study of dynamical fractional wormholes, rotating configurations, generalized matter sources, and extensions to modified gravitational theories and cosmological backgrounds.



\section*{Acknowledgments}
The author is grateful to Arshiya Farhath G D, Binaya K Baral, Arpit Mohapatra, and Nirupama Rauto for valuable discussions and assistance during the completion of this work. Special thanks are also extended to reviewers for their insightful feedback and helpful recommendations.

\section*{Declarations}

\subsection*{Author contribution}
The author performed all tasks related to this study, including research conceptualization, methodology, data analysis, manuscript drafting, and final review.

\subsection*{Funding}
No funding was received to assist with the preparation of this manuscript.

\subsection*{Conflict of Interest}
The authors declare no conflict of interest.

\subsection*{Data Availability Statement}
Data sharing not applicable to this article as no datasets were generated or analysed during the current study.

\section*{Keywords}
Super-Hyperbolic Wormhole, Einstein Field Equations, Traversable wormhole, Catenoid, Fractional curvature spacetime

\nocite{*}

\bibliography{apssamp}


\appendix

\section{Choice of the Redshift Function}

This appendix collects the derivation and physical interpretation of the redshift profile used in Section~\ref{sec:level4}. Throughout this appendix, the metric convention adopted in the manuscript is
\[
ds^{2}=-e^{2\Phi(r)}dt^{2}+\cdots,
\]
so that $e^{\Phi(r)}$ is the lapse function and $e^{2\Phi(r)}$ is the corresponding metric coefficient in the time sector.
To model a smooth and throat-symmetric tidal sector, we choose the redshift function
\begin{equation*}
\Phi(r)
=
\Phi_{0}\,
\ln\!\left[
\cosh\!\left(\frac{r-r_{\mathrm{th}}}{\Phi_{1}}\right)
\right].
\end{equation*}
The throat position is denoted by $r_{\mathrm{th}}$. This profile is even about the throat, analytic for all finite $r$, and normalized so that
\[
\Phi(r_{\mathrm{th}})=0.
\]
Hence the time component of the metric satisfies
\[
g_{tt}(r_{\mathrm{th}})=-1,
\]
so the throat is used as the local reference level for the redshift.
The corresponding lapse is
\begin{equation}
e^{2\Phi(r)}
=
\cosh^{2\Phi_{0}}\!\left(\frac{r-r_{\mathrm{th}}}{\Phi_{1}}\right).
\end{equation}

The throat radius, $r_{th}$, represents the location of the wormhole's throat - the narrowest point of the tunnel connecting two regions of spacetime. At the throat, the potential is zero. This is the reference point for the energy of the system. The dimensionless redshift amplitude parameter, $\Phi_{0}$, controls the magnitude or strength of the gravitational field. A larger $\Phi_{0}$ means the time dilation effects are more extreme. If one was an observer sitting at a distance from the throat, a higher $\Phi_{0}$ would mean one's clock runs significantly differently compared to a clock at the throat. The characteristic length scale factor, $\Phi_{1}$, determines the steepness or width of the gravitational potential. For a small $\Phi_{1}$, the potential changes very rapidly as one moves away from the throat, called as steep gravity, whereas a large $\Phi_{1}$, the potential varies slowly, called gentle gravity. It sets the size of the region where the gravitational effects are the most significant.

Since $\cosh x\geq 1$ for all real $x$, the metric coefficient $e^{2\Phi(r)}$, known as the Redshift Factor, is strictly positive and finite for every finite $r$. Thus the redshift profile itself does not generate a horizon inside the domain where it is used, allowing a traveler to pass through the throat safely (hence, a traversable wormhole). It determines how time dilation affects an observer at position $r$ relative to the throat. An observer floating exactly at the throat would feel zero gravitational force. The gravity would pull them back towards the throat if they moved away in either direction. For very large distances $(r \rightarrow \infty),$ the term $\ln(\cosh (r))$ behaves like linear $r$. This implies the gravitational potential grows indefinitely rather than fading to zero (like Earth's gravity does) and the equation likely describes only the interior geometry (near the throat) or a universe that is not empty at infinity (like an Anti-de Sitter space).

\end{document}